\documentclass[preprint,12pt]{elsarticle}

\usepackage{graphicx}

\usepackage{amssymb}
\usepackage{multirow}
\usepackage{subfig}
\biboptions{sort&compress}

\newlength{\dslashwidth}

\newcommand{\beq}{\begin{equation}} 
\newcommand{\eeq}{\end{equation}}
\newcommand{\beqa}{\begin{eqnarray}} 
\newcommand{\eeqa}{\end{eqnarray}}
\newcommand{\newc}{\newcommand}
\newcommand{\bq}{\begin{equation}}
\newcommand{\eq}{\end{equation}}
\newcommand{\ba}{\begin{array}}
\newcommand{\ea}{\end{array}}
\newcommand{\bqa}{\begin{eqnarray}}
\newcommand{\eqa}{\end{eqnarray}}

\newcommand{\lnf}{{\ifmmode \Lambda^{(N_f)} \else $\Lambda^{(N_f)}$\fi}}
\newcommand{\ms}{{\ifmmode \overline{MS} \else $\overline{MS}$\fi}}
\newcommand{\dr}{{\ifmmode \overline{DR} \else $\overline{DR}$\fi}}
\newcommand{\lms}{{\ifmmode \Lambda^{(5)}_{\overline{MS}} \else $\Lambda^{(5)}_{\overline{MS}}$\fi}}
\newcommand{\lam}{{\ifmmode \Lambda \else $\Lambda$\fi}}
\newcommand{\mev}{{\ifmmode {\rm MeV} \else ${\rm MeV}$\fi}}
\newcommand{\gev}{{\ifmmode {\rm GeV} \else ${\rm GeV}$\fi}}
\newcommand{\gevc}{{\ifmmode {\rm GeV/c^2} \else ${\rm GeV/c^2}$\fi}}
\newcommand{\tev}{{\ifmmode {\rm TeV} \else ${\rm TeV}$\fi}}
\newcommand{\tevc}{{\ifmmode {\rm TeV/c^2} \else ${\rm TeV/c^2}$\fi}}
\newcommand{\lp}{{\ifmmode L^+  \else $L^+$\fi}}
\newcommand{\lm}{{\ifmmode L^-  \else $L^-$\fi}}
\newcommand{\mlp}{{\ifmmode M(L^-) \else $M(L^-)$\fi}}
\newcommand{\mlz}{{\ifmmode M(L^0) \else $M(L^0)$\fi}}
\newcommand{\lz}{{\ifmmode L^0 \else $L^0$\fi}}
\newcommand{\ev}{{\ifmmode GeV/c^2 \else $GeV/c^2$\fi}}
\newcommand{\tri}{{\ifmmode \triangleup \else $\triangleup$\fi}}
\newcommand{\unl}{{\ifmmode U_{lL^0} \else $U_{lL^0}$\fi}}\newcommand{\gL}{{\ifmmode g_L \else $g_{L}$\fi}}
\newcommand{\gR}{{\ifmmode g_R  \else $g_{R}$\fi}}
\newcommand{\gumu}{{\ifmmode \gamma^{\mu} \else $\gamma^{\mu}$\fi}}
\newcommand{\gunu}{{\ifmmode \gamma^{\nu} \else $\gamma^{\nu}$\fi}}
\newcommand{\gdmu}{{\ifmmode \gamma_{\mu} \else $\gamma_{\mu}$\fi}}
\newcommand{\gdnu}{{\ifmmode \gamma_{\nu} \else $\gamma_{\nu}$\fi}}
\newcommand{\stw}{{\ifmmode\sin^2\theta_W \else $\sin^{2}\theta_{W}$ \fi}}
\newcommand{\sws}{{\ifmmode \;\sin^2\theta_W  \else $\;\sin^{2}\theta_{W}$ \fi}}
\newcommand{\cws}{{\ifmmode \;\cos^2\theta_W  \else $\;\cos^{2}\theta_{W}$ \fi}}
\newcommand{\sw}{{\ifmmode \;\sin\theta_W  \else $\sin\theta_{W}$ \fi}}
\newcommand{\cw}{{\ifmmode \;\cos\theta_W  \else $\;\cos\theta_{W}$ \fi}}
\newcommand{\tw}{{\ifmmode \;\tan\theta_W  \else $\;\tan\theta_{W}$ \fi}}
\newcommand{\qq}{{\ifmmode q\overline{q} \else $q\overline{q}$\fi}}
\newcommand{\lR}{{\ifmmode l_R \else $l_R$\fi}}
\newcommand{\lL}{{\ifmmode l_L \else $l_L$\fi}}
\newcommand{\nt}{{\ifmmode \nu_{\tau} \else $\nu_{\tau}$\fi}}
\newcommand{\nuR}{{\ifmmode \nu_R  \else $\nu_R$\fi}}
\newcommand{\nuL}{{\ifmmode \nu_L  \else $\nu_L$\fi}}
\newcommand{\qR}{{\ifmmode g_R  \else $q_R$\fi}}
\newcommand{\qL}{{\ifmmode q_L  \else $q_L$\fi}}
\newcommand{\qRp}{{\ifmmode q_R'  \else $q_{R}$'\fi}}
\newcommand{\qLp}{{\ifmmode q_L'  \else $q_{L}$'\fi}}
\newcommand{\est}{{\ifmmode e^{\bf \ast} \else $e^{\bf \ast}$\fi}}
\newcommand{\lst}{{\ifmmode l^{\bf \ast} \else $l^{\bf \ast}$\fi}}
\newcommand{\must}{{\ifmmode \mu^{\bf \ast} \else $\mu^{\bf \ast}$\fi}}
\newcommand{\taust}{{\ifmmode \tau^{\bf \ast} \else $\tau^{\bf \ast}$ \fi}}
\newcommand{\pperp}{{\ifmmode p_t  \else $p_t$\fi}}
\newcommand{\et}{{\ifmmode E_t  \else $E_t$\fi}}
\newcommand{\xt}{{\ifmmode x_t  \else $x_t$\fi}}
\newcommand{\smumu}{{\ifmmode \sigma_{\mu\mu}  \else $\sigma_{\mu\mu}$ \fi}}
\newcommand{\eg}{{\ifmmode e\gamma  \else $e\gamma$\fi}}
\newcommand{\epem}{{\ifmmode e^+e^-  \else $e^+e^-$\fi}}
\newcommand{\lplm}{{\ifmmode L^+L^-  \else $L^+L^-$\fi}}
\newcommand{\pp}{{\ifmmode p\overline p  \else $p\overline p$\fi}}
\newcommand{\llz}{{\ifmmode L^0\overline{L}^0 \else $L^0\overline{L}^0$\fi}}
\newcommand{\epemt}{{\ifmmode e^+e^- \to  \else $e^+e^- \to$\fi}}
\newcommand{\eb}{{\ifmmode E_{beam}  \else $E_{beam}$\fi}}
\newcommand{\ip}{{\ifmmode pb^{-1}  \else $pb^{-1}$\fi}}
\newcommand{\upm}{{\ifmmode ^{\pm}  \else $^{\pm}$\fi}}
\newcommand{\de}{{\ifmmode ^{\circ}  \else $^{\circ}$ \fi}}
\newcommand{\appr}{{\ifmmode \sim \else $\sim$ \fi}}
\newcommand{\corresp}{{\ifmmode \stackrel{\wedge}{=} \else $\stackrel{\wedge}{=}$ \fi}}
\newcommand{\sqrts}{{\ifmmode \sqrt{s} \else $\sqrt{s}$\fi}}
\newcommand{\zz}{{\ifmmode Z^0  \else $Z^0$\fi}}
\newcommand{\mz}{{\ifmmode M_{Z}  \else $M_{Z}$\fi}}
\newcommand{\mzs}{{\ifmmode M_{Z}^2  \else $M_{Z}^2$\fi}}
\newcommand{\mw}{{\ifmmode M_{W}  \else $M_{W}$\fi}}
\newcommand{\mws}{{\ifmmode M_{W}^2  \else $M_{W}^2$\fi}}
\newcommand{\mh}{{\ifmmode M_{Higgs}  \else $M_{Higgs}$\fi}}
\newcommand{\msusy}{{\ifmmode M_{SUSY}  \else $M_{SUSY}$\fi}}
\newcommand{\msusys}{{\ifmmode M_{SUSY}^2  \else $M_{SUSY}^2$\fi}}
\newcommand{\su}{{\ifmmode SU(3)_C\otimes\- SU(2)_L\otimes\- U(1)_Y \else $SU(3)_C\otimes\A0SU(2)_L\otimes U(1)_Y$\fi}}
\newcommand{\suthree}{{\ifmmode SU(3)_C  \else $SU(3)_C$\fi}}
\newcommand{\sutwo}{{\ifmmode  SU(2)_L\otimes U(1)_Y \else $SU(2)_L\otimes U(1)_Y$\fi}}
\newcommand{\taup}{{\ifmmode \tau_{proton} \else $\tau_{proton}$\fi}}
\newcommand{\as}{{\ifmmode \alpha_{s}  \else $\alpha_{s}$\fi}}
\newcommand{\mgut}{{\ifmmode M_{GUT}  \else $M_{GUT}$\fi}}
\newcommand{\mguts}{{\ifmmode M_{GUT}^2  \else $M_{GUT}^2$\fi}}
\newcommand{\mzero}{{\ifmmode m_0        \else $m_0$\fi}}
\newcommand{\mhalf}{{\ifmmode m_{1/2}    \else $m_{1/2}$\fi}}
\newcommand{\sq}{{\ifmmode \tilde{q}    \else $\tilde{q}$\fi}}
\newcommand{\gl}{{\ifmmode \tilde{g}    \else $\tilde{g}$\fi}}
\newcommand{\mb}{{\ifmmode m_{b}    \else $m_{b}$\fi}}
\newcommand{\mt}{{\ifmmode m_{t}    \else $m_{t}$\fi}}
\newcommand{\mts}{{\ifmmode m_{t}^2    \else $m_{t}^2$\fi}}

\newcommand{\mtau}{{\ifmmode m_{\tau}  \else $m_{\tau}$\fi}}
\newcommand{\dpp}{{\ifmmode \delta_{pert} \else $\delta_{pert}$\fi}}
\newcommand{\dnp}{{\ifmmode\delta_{non-pert}\else$\delta_{non-pert}$\fi}}
\newcommand{\dew}{{\ifmmode \delta_{\rm EW}\else $\delta_{\rm EW}$\fi}}
\newcommand{\rt}{{\ifmmode R_{\tau}  \else $R_{\tau} $\fi}}
\newcommand{\rz}{{\ifmmode R_{Z}  \else $R_{Z} $\fi}}

\newcommand{\swb}{{\ifmmode \sin^2\theta_{\overline{MS}} \else $\sin^2\theta_{\overline{MS}}$\fi}}
\newcommand{\cwb}{{\ifmmode \cos^2\theta_{\overline{MS}} \else $\cos^2\theta_{\overline{MS}}$\fi}}

\newc\AIPCP[3] {{\em AIP Conf. Proc.} {\bf #1} (#2) #3}
\newc\AJ[3] {{\em Astrophys. J.} {\bf #1} (#2) #3}
\newc\AMS[3] {{\em Ann. Math. Statist.} {\bf #1} (#2) #3}
\newc\AP[3] {{\em Ann. Phys.} {\bf #1} (#2) #3}
\newc\APJ[3] {{\em Astropart. J.} {\bf #1} (#2) #3}
\newc\APP[3] {{\em Astropart. Phys.} {\bf #1} (#2) #3}
\newc\APS[3] {{\em Astrophys. J. Suppl.} {\bf #1} (#2) #3}
\newc\ARNPS[3] {{\em Ann. Rev. Nucl. Part. Sci.} {\bf C#1} (#2) #3}
\newc\BA[3] {{\em Bayesian Anal.} {\bf C#1} (#2) #3}
\newc\CPC[3] {{\em Comput. Phys. Commun.} {\bf C#1} (#2) #3}
\newc\CP[3] {{\em Contemp. Phys.} {\bf #1} (#2) #3}
\newc\EPJ[3] {{\em Euro. Phys. Journ.} {\bf C#1} (#2) #3}
\newc\JCAP[3] {{\em JCAP} {\bf #1} (#2) #3}
\newc\JHEP[3] {{\em JHEP} {\bf #1} (#2) #3}
\newc\JPG[3] {{\em J. Phys.} {\bf G #1} (#2) #3}
\newc\IJMP[3] {{\em Int. J. Mod. Phys.} {\bf A #1} (#2) #3}
\newc\MNRAS[3] {{\em Mon. Not. Roy. Astron. Soc.} {\bf #1} (#2) #3}
\newc\MPL[3] {{\em Mod. Phys. Lett.} {\bf A #1} (#2) #3}
\newc\NAR[3] {{\em New Astron. Rev.} {\bf #1} (#2) #3}
\newc\NCA[3] {{\em Nuovo Cimento} {\bf #1} (#2) #3}
\newc\NIM[3] {{\em Nucl. Instrum. Methods} {\bf #1} (#2) #3}
\newc\NIMA[3] {{\em Nucl. Instrum. Methods} {\bf A #1} (#2) #3}
\newc\NAT[3] {{\em Nature} {\bf #1} (#2) #3}
\newc\NPB[3] {{\em Nucl. Phys.} {\bf B #1} (#2) #3}
\newc\NPA[3] {{\em Nucl. Phys.} {\bf A #1} (#2) #3}
\newc\NPPS[3] {{\em Nucl. Phys. Proc. Suppl.} {\bf #1} (#2) #3}
\newc\PLB[3] {{\em Phys. Lett.} {\bf B #1} (#2) #3}
\newc\PR[3] {{\em Phys. Rep.} {\bf #1} (#2) #3}
\newc\PRL[3] {{\em Phys. Rev. Lett.} {\bf #1} (#2) #3}
\newc\PRD[3] {{\em Phys. Rev.} {\bf D #1} (#2) #3}
\newc\PRC[3] {{\em Phys. Rev.} {\bf C #1} (#2) #3}
\newc\PTP[3] {{\em Prog. Theor. Phys.} {\bf #1} (#2) #3}
\newc\RMP[3] {{\em Rev. Mod. Phys.} {\bf #1} (#2) #3 }
\newc\RPP[3] {{\em Rept. Prog. Phys.} {\bf #1} (#2) #3 }
\newc\SC[3] {{\em Science} {\bf #1} (#2) #3 }
\newc\ZPC[3] {{\em Z. Phys.} {\bf C #1} (#2) #3}
\newc\Err[3] {{\em Erratum-ibid.} {\bf #1} (#2) #3 }

\journal{Physics Letters B}

\begin{document}

\begin{frontmatter}


%
\title{Higgs Branching Ratios in Constrained Minimal and Next-to-Minimal Supersymmetry Scenarios Surveyed}

\author[label1]{C. Beskidt}\ead{conny.beskidt@kit.edu}
\author[label1]{W. de Boer}\ead{wim.de.boer@kit.edu}
\author[label1,label2]{D.I. Kazakov}
\author[label1]{Stefan Wayand}
\address[label1]{Institut f\"ur Experimentelle Kernphysik, Karlsruhe Institute of Technology, P.O. Box 6980, 76128 Karlsruhe, Germany}
\address[label2]{Bogoliubov Laboratory of Theoretical Physics, Joint Institute for Nuclear Research, 141980, 6 Joliot-Curie, Dubna, Moscow Region, Russia}

\begin{abstract}
In the CMSSM the heaviest scalar and pseudo-scalar Higgs bosons decay largely into b-quarks and tau-leptons because of the large $\tan\beta$ values favored by the relic density. In the NMSSM the number of possible decay modes is much richer. In addition to the CMSSM-like scenarios, the decay of the heavy Higgs bosons is preferentially into top quark pairs (if kinematically allowed),  lighter Higgs bosons or neutralinos, leading to invisible decays.
We provide a scan over the NMSSM parameter space to project the 6D parameter space of the Higgs sector on the 3D space of the Higgs masses to determine the range of branching ratios as function of the Higgs boson mass for all Higgs bosons. Specific LHC benchmark points are proposed, which represent the salient NMSSM features.

\end{abstract}

\end{frontmatter}


\section{Introduction}
\label{Introduction}

A light Higgs boson below 135 GeV is predicted within Supersymmetry (SUSY)\cite{Haber:1984rc,deBoer:1994dg,Martin:1997ns}. So the discovery of a Higgs-like boson with a mass of 125 GeV \cite{Aad:2012tfa,Chatrchyan:2012ufa} strongly supports SUSY although no SUSY particles have been found so far. The precise value of the Higgs mass depends on radiative corrections.  Within the constrained minimal supersymmetric standard model (CMSSM) \cite{Kane:1993td} the tree level Higgs boson mass is below the $Z^0$-boson mass $M_Z$ (91 GeV) and to reach the observed mass of 125 GeV 
the radiative corrections from stop loops have to be large, see e.g. \cite{Beskidt:2012sk,Buchmueller:2013rsa,Fowlie:2012im,Bechtle:2013mda} and references therein. 
However, a 125 GeV Higgs boson is easily obtained in the minimal extension of the CMSSM where an additional Higgs singlet is introduced, since then the tree level value of the Higgs boson can be above $M_Z$. The reason is simple: within the so-called next-to-minimal supersymmetric standard model (NMSSM) \cite{Ellwanger:2009dp} the mixing with the additional Higgs singlet increases the Higgs mass at tree level \cite{King:2012tr,Cao:2012fz,Belanger:2012tt,Ellwanger:2012ke,King:2012is,Vasquez:2012hn,Beskidt:2013gia,Badziak:2013bda}, so the radiative corrections from the stop loops do not need multi-TeV stop squarks in the NMSSM, thus avoiding  the fine-tuning problem \cite{Martin:1997ns,deBoer:1994dg,Haber:1984rc}.
The addition of a Higgs singlet yields more parameters in the Higgs sector to cope with the interactions between the singlet and the doublets and the singlet self interaction. Furthermore, the supersymmetric partner of the singlet leads to an additional Higgsino, thus extending the neutralino sector from 4 to 5 neutralinos.
These additional particles and their interactions lead to a large parameter space, even if one considers the well-motivated subspace with unified masses and couplings at the GUT scale. 

On the other hand, experiments are mostly interested in possible ranges of Higgs masses and branching ratios.
With 5 neutral Higgs masses, of which one has to be 125 GeV and  two of the heavy neutral Higgses masses are practically mass-degenerate, one is left with a 3-dimensional (3D) space in the Higgs masses in contrast to the 6-dimensional  (6D) parameter space of the constrained $Z_3$-invariant NMSSM Higgs sector.
A certain point in the Higgs mass space can be obtained for several combinations of the 6D parameter space, which in turn leads to a range of branching ratios of the Higgs bosons.

In this paper we ventured to project the 6D parameter space on the 3D  space of Higgs masses to obtain the expected range of branching ratios as function of the Higgs mass for each Higgs boson.
This allows us to look for the distinctive features between the NMSSM and CMSSM.
 After a short summary of the Higgs and gaugino sectors in the CMSSM and NMSSM we discuss the fit strategy to project the 6D parameter space on the 3D neutral Higgs mass space. We conclude by summarizing the branching ratios of both models and selected benchmark points showing the salient features of the NMSSM. 

\section{NMSSM Higgs sector}
\label{higgs}
We focus on the well-motivated  semi-constrained next-to-minimal supersymmetric standard model (NMSSM), as described in Ref. \cite{Ellwanger:2009dp} and  use the corresponding code NMSSMTools 4.6.0 \cite{Das:2011dg} to calculate the SUSY mass spectrum, Higgs boson masses and  branching ratios from the NMSSM parameters.

 Within the NMSSM the Higgs fields consist of the two Higgs doublets  ($H_u, H_d$), which appear in the MSSM as well, but together with
an additional complex Higgs singlet $S$. 
In addition, we have the GUT scale parameters of the CMSSM: $\mzero$, $\mhalf$ and $A_0$, where $\mzero$($\mhalf$) are the common mass scales of the spin 0(1/2) SUSY particles at the GUT scale and $A_0$ is the trilinear coupling of the CMSSM Higgs sector at the GUT scale. In total the semi-constrained NMSSM has nine free parameters:
\begin{equation}
 \mzero,~ \mhalf,~A_0,~ \tan\beta,  ~ \lambda, ~\kappa,  ~A_\lambda, ~A_\kappa, ~\mu_{eff}.
\label{params}
\end{equation}
Here $\tan\beta$ corresponds to the ratio of the vevs of the Higgs doublets, i.e. $\tan\beta\equiv v_u/v_d$,
$ \lambda$ represents the coupling between the Higgs singlet and doublets ($\lambda S H_u\cdot  H_d$),  $~\kappa$ the self-coupling of the singlet ($\kappa S^3/3$);   $A_\lambda$ and $A_\kappa$ are the corresponding trilinear soft breaking terms, $\mu_{eff}$ represents an effective Higgs mixing parameter and is related to the vev of the singlet $s$ via the coupling $\lambda$, i.e. $\mu_{eff}\equiv\lambda s$. Therefore, $\mu_{eff}$ is naturally of the order of the electroweak scale, thus avoiding the $\mu$-problem \cite{Ellwanger:2009dp}. The latter six parameters in Eq. \ref{params} form the 6D parameter space of the NMSSM Higgs sector.

The neutral components from the two Higgs doublets and singlet  mix to form three physical CP-even
 scaler ($S$)  bosons and two physical CP-odd pseudo-scalar ($P$)  bosons.

The elements of the corresponding mass matrices at
tree level  read \cite{Miller:2003ay}:
\begin{eqnarray}
{\cal M}^2_{S,11}&=&M_A^2+(M_Z^2-\lambda^2v^2)\sin^22\beta,\nonumber\\
{\cal M}^2_{S,12}&=&-\frac{1}{2}(M_Z^2-\lambda^2v^2)\sin4\beta,\nonumber\\
{\cal M}^2_{S,13}&=&-\frac{1}{2}(M_A^2\sin2\beta+\frac{2\kappa\mu^2}{\lambda})\frac{\lambda v}{\mu}\cos2\beta,\nonumber\\
{\cal M}^2_{S,22}&=&M_Z^2\cos^22\beta+\lambda^2v^2\sin^22\beta,\label{mix}\\
{\cal M}^2_{S,23}&=& 2 \lambda \mu v \left[1 - (\frac{M_A \sin 2\beta}{2 \mu} )^2
-\frac{\kappa}{2 \lambda}\sin2\beta\right],\nonumber\\
{\cal M}^2_{S,33}&=& \frac{1}{4} \lambda^2 v^2 (\frac{M_A \sin 2\beta}{\mu})^2
+ \frac{\kappa\mu}{\lambda} (A_\kappa +  \frac{4\kappa\mu}{\lambda} )
 - \frac{1}{2} \lambda \kappa v^2 \sin 2 \beta,\nonumber 
\end{eqnarray}

\begin{eqnarray}
{\cal M}^2_{P,11}&=&\frac{\mu (\sqrt{2}A_\lambda+\kappa \frac{\mu}{\lambda})}{\sin 2 \beta}=M^2_A,\nonumber\\
{\cal M}^2_{P,12}&=&\frac{1}{\sqrt{2}}\left( M^2_A \sin 2 \beta - 3 \frac{\kappa}{\lambda}\mu^2 \right)\frac{v\lambda}{\mu},\\
{\cal M}^2_{P,22}&=&\frac{1}{2}\left( M^2_A \sin 2 \beta + 3 \frac{\kappa}{\lambda} \mu^2 \right ) \frac{v^2}{\mu^2}\lambda^2 \sin 2 \beta - \frac{3}{\sqrt{2} \frac{\kappa}{\lambda}\mu A_\kappa}.\nonumber
\end{eqnarray}

One observes that the element ${\cal M}^2_{S,22}$, which corresponds to the tree-level term of the lightest CMSSM Higgs boson, can be above $M^2_Z$ because of the $\lambda^2v^2\sin^2 2 \beta$ term.
The diagonal element ${\cal M}^2_{P,11}$ at tree level corresponds to the pseudo-scalar Higgs bosons in the MSSM limit of small $\lambda$, so it is called $M_A$. 
   $M^2_{S,33}$ and $M^2_{P,22}$ correspond to the diagonal terms for the additional scalar and pseudo-scalar Higgs boson not present in the MSSM. 
The mass of the heaviest scalar and pseudo-scalar Higgs bosons  are usually close to each other, since the dominant term at tree level is in both cases $M_A^2$, as can be seen from a comparison of $ {\cal M}^2_{S,11}$ and  ${\cal M}^2_{P,11}$.
The mass eigenstate of the charged Higgs fields reads:

\begin{eqnarray}
M^2_{H^\pm}&=& M^2_A +M^2_W-\frac{1}{2}(\lambda v)^2.
\end{eqnarray}

Note that the heavy charged and heavy neutral Higgs masses are  all of the order of $M_A$ and  largely independent of the SUSY masses.

\section{CMSSM and NMSSM gaugino sector}
\label{neu-char-sector}

Within the NMSSM the singlino, the superpartner of the Higgs singlet, mixes with the gauginos and Higgsinos, leading to an additional fifth neutralino. The resulting mixing matrix reads \cite{Ellwanger:2009dp,Staub:2010ty}:

\beq\label{eq1}
{\cal M}_0 =
\left( \ba{ccccc}
M_1 & 0 & -\frac{g_1 v_d}{\sqrt{2}} & \frac{g_1 v_u}{\sqrt{2}} & 0 \\
0 & M_2 & \frac{g_2 v_d}{\sqrt{2}} & -\frac{g_2 v_u}{\sqrt{2}} & 0 \\
-\frac{g_1 v_d}{\sqrt{2}} & \frac{g_2 v_d}{\sqrt{2}} & 0 & -\mu_\mathrm{eff} & -\lambda v_u \\
\frac{g_1 v_u}{\sqrt{2}} & -\frac{g_2 v_u}{\sqrt{2}}& -\mu_\mathrm{eff}& 0 & -\lambda v_d \\
0& 0& -\lambda v_u&  -\lambda v_d & 2 \kappa s 
\ea \right)
\eeq
with the gaugino masses $M_1$, $M_2$, the gauge couplings $g_1$, $g_2$ and the Higgs mixing parameter $\mu_{eff}$ as parameters.
Furthermore, the vacuum expectation values of the two Higgs doublets $v_d$,$v_u$, the singlet s and the Higgs couplings $\lambda-\kappa$ enter the neutralino mass matrix. 
The upper $4 \times 4$ submatrix of the neutralino mixing matrix corresponds to the MSSM neutralino mass matrix, see e.g. Ref. \cite{Martin:1997ns}. 
Since the additional Higgs singlino affects only the neutral gaugino sector, the mixing matrix for the charginos in the NMSSM and CMSSM are identical: 

\beq\label{eq2}
{\cal M}_\pm =
\left( \ba{cc}
M_2 & g_2 v_u\\
g_2 v_d & \mu_{eff}\\
\ea \right).
\eeq

To obtain the mass eigenstates the mass matrices have to be diagonalized. Typically the diagonal elements in Eq. \ref{eq1} and \ref{eq2} dominate over the off-diagonal terms, so the neutralino masses are of the order of $M_1$, $M_2$, the Higgs mixing parameter $\mu_{eff}$ and in case of the NMSSM $2 \kappa/ \lambda \mu_{eff} $. The chargino masses are of the order of $M_2$ and $\mu_{eff}$.

Since we use GUT scale input parameters and the mass spectrum at the low mass SUSY scales is calculated via the renormalization group equations (RGEs), the masses are correlated. The gaugino masses are proportional to $m_{1/2}$ \cite{Martin:1997ns,Haber:1984rc,deBoer:1994dg}:
\beq\label{eq3}
M_1\approx 0.4 m_{1/2},~ 
M_2\approx 0.8 m_{1/2},~
M_3\approx M_{\tilde{g}} \approx 2.7 m_{1/2}.  
\eeq 

This leads to bino-like light neutralinos and wino-like light charginos in the CMSSM, since $\mu$ is typically much larger than $m_{1/2}$ to fulfill radiative electroweak symmetry breaking (EWSB) \cite{Haber:1984rc,deBoer:1994dg,Martin:1997ns}.
In the NMSSM $\mu_{eff}$ is an input parameter and it can be chosen such that it is of the order of the electroweak scale. This changes both the neutralino and chargino sector. In such natural NMSSM scenarios the lightest neutralino is singlino-like and its mass can be degenerate to the second/third neutralino and the lightest chargino, which all have a mass of the order of $\mu_{eff}$.

\section{Analysis}
\label{analysis}

As discussed in sect. \ref{higgs} the number of free parameters in the NMSSM increases with respect to the MSSM. Six of the nine free NMSSM parameters enter the Higgs sector. 
For each set of parameters the Higgs boson masses are completely determined:  3 scalar Higgs masses $m_{H_i}$, 2 pseudo-scalar Higgs masses $m_{A_i}$ and the charged Higgs boson mass $m_{H^\pm}$. The index $i$ increases with increasing mass. The masses of $A_2$, $H_3$ and $H^\pm$ are of the order of $M_A$, if $M_A >> M_Z$. Then only one of the masses  is needed. Furthermore, either $H_1$ or $H_2$ has to be the observed Higgs boson with a mass of 125 GeV, so there are only 3 free neutral Higgs boson masses in the NMSSM, i.e. a 3D parameter space, e.g $m_{A_1}$, $m_{H_1}$ and $m_{H_3}\approx m_{A_2} \approx m_{H^\pm}$.
Instead of scanning over the 6D parameter space of the couplings to determine the range of Higgs boson masses, which was done by other groups in the MSSM \cite{Arganda:2012qp,Arganda:2013ve}, one can invert the problem and scan  the 3D parameter space of the Higgs boson masses and check which region of the 6D parameter space leads to a given point in the 3D Higgs mass space.

We proceed as follows: we divide the $m_{H_1}-m_{H_3}$ mass plane in a grid with fine mass bins for a certain value of $m_{A_1}$. These grids were repeated with
the values of $m_{A_1}$ varying between 25 and 500 GeV, while $m_{H_1}$ ranges from 5 to 125 GeV in steps of 5 GeV. The heavy Higgs boson mass $m_{H_3}$ was allowed to vary between 100 GeV and 2 TeV.

For each bin in each grid for a given $m_{A_1}$ one can use Minuit \cite{James:1975dr} to determine the corresponding NMSSM parameters at the GUT scale using a $\chi^2$ function, which reads: 

\beq\label{eq5}
\chi^2_{tot}=\chi^2_{H_1}+\chi^2_{H_2}+\chi^2_{H_3}+\chi^2_{LEP}.
\eeq 

The $\chi^2$ contributions are 

\begin{itemize} 
\item $\chi^2_{H_1}=(m_{H_1} - m_{grid,H_1})^2/\sigma^2_{H_1}$. This term requires the NMSSM parameters to be adjusted such that  the mass of the lightest Higgs boson mass $m_{H_1}$ agrees with the chosen point in the 3D mass space $m_{grid,H_1}$. $m_{H_1}$  has always a mass below 
 the observed Higgs boson mass.  The value of $\sigma^2_{H_1}$ is set to 2 GeV. 
\item $\chi^2_{H_2}=(m_{H_2} - m_{obs})^2/\sigma^2_{SM}+\sum_i (c^i_{H_2} - c_{obs})^2/\sigma^2_{coup}$: since the lightest Higgs boson $H_1$ has a mass below 125 GeV, the second lightest Higgs boson has to represent the observed Higgs boson with couplings close to the SM couplings, as required by the last term.  $c^i_{H_2}$ represents the reduced couplings of $H_2$  which is the ratio of the coupling of $H_2$ to particle $i=f_u,f_d,W/Z,\gamma$ divided by the SM coupling. The observed couplings $c_{obs}$ agree within 10\% with the SM couplings, so $\sigma^2_{coup}=0.1$. The first term is analogous to the term for  $m_{H_1}$, except that the mass of the second lightest Higgs boson should have the observed Higgs boson mass, so $m_{obs}$ is set to $125.2$ GeV. The corresponding uncertainty $\sigma^2_{SM}$ equals $1.9$ GeV and results from the linear addition of the experimental and theoretical ($1.5$ GeV) uncertainties. 
\item $\chi^2_{H_3}=(m_{H_3} - m_{grid,H_3})^2/\sigma^2_{H_3}$: as $\chi^2_{H_1}$, but for the heavy scalar Higgs boson $H_3$.
\item $\chi^2_{LEP}$: includes the LEP constraints on the couplings of a light Higgs boson below 115 GeV and the limit on the chargino mass as discussed in Ref. \cite{Beskidt:2014kon}.      
\end{itemize}

We allowed also the rare cases, where the lightest Higgs boson is the observed Higgs boson with SM-like couplings and  $m_{H_2}$ is above the observed mass (usually slightly).
 In addition, we checked what happens if one adds the cosmological constraints assuming the LSP (largely singlino) provides the relic density and gives a nucleon scattering cross section consistent with the direct DM searches. These dark matter constraints are calculated with micrOmegas \cite{Belanger:2010pz}, as interfaced within NMSSMTools. 

In summary, the analysis looks like one has observed all Higgs boson masses and tries to infer the corresponding region of the 6D NMSSM para\-meter space
with the option to include the cosmological constraints. From the allowed region of couplings in the 6D space one can then deduce the allowed range of branching ratios for the considered Higgs boson masses in the 3D mass space.

The determination of the 6D parameter set to obtain a certain Higgs mass combination is not unique, as can be easily seen already from the approximate expression for the 125 GeV Higgs boson \cite{Ellwanger:2009dp}: 
\beq\label{eq4}
M_{H}^2\approx M_Z^2\cos^2 2\beta+ \Delta_{\tilde{t}} + \lambda^2 v^2 \sin^2 2\beta - \frac{\lambda^2}{\kappa^2}(\lambda-\kappa \sin 2 \beta)^2.  
\eeq 
The first two terms are identical to the CMSSM, where the first tree level term can become as large as  $M_Z^2$ for large $\tan\beta$, but in the CMSSM the difference between  $M_Z$  and 125 GeV  has to originate mainly from the logarithmic stop mass corrections $\Delta_{\tilde{t}}$. The two remaining terms originate from the mixing with the singlet of the NMSSM and become large for large values of
 the couplings $\lambda$ and $\kappa$ and small  $\tan\beta$.  This is what we call \textit{scenario I}.  However, the 125 GeV Higgs boson mass can also be reached by a trade-off between the first two CMSSM terms and last two NMSSM terms using smaller couplings and larger $\tan\beta$ values. This is what we call \textit{scenario II}. 
These scenarios have distinctly different signatures. In \textit{scenario II} the decays of the heavy Higgs bosons to down type fermions are enhanced by $\tan^2\beta$, thus preferring decays to  b-quarks and $\tau$ leptons, while decays to top quarks are suppressed by $1/\tan^2\beta$. In \textit{scenario I}, the large values of the couplings $\lambda-\kappa$  lead to decays of the heaviest scalar Higgs boson to the two lighter ones which is dominant for heavy Higgs boson masses below the $t\bar{t}$ decay threshold of about 400 GeV. For $m_{H_3} > 400$ GeV the decay into $t\bar{t}$ starts to dominate. 
\begin{table}[!t]
\small
\centering
\caption{ The two main NMSSM scenarios corresponding to different ranges of the masses and couplings which are associated with different numbered benchmark points (BMP). The range of $\tan\beta$ is determined by the observed Higgs mass for a given range of the couplings $\kappa$ and $\lambda$. 
\label{t1}}
\begin{tabular}{l|c|c}
	\hline\noalign{\smallskip}
	scenario & I & II \\
	\noalign{\smallskip}\hline\noalign{\smallskip}
	\multirow{2}*{couplings}	 & $\tan\beta < 10 $ & $\tan\beta > 10$ \\
	 & $\lambda,\kappa$ large & $\lambda,\kappa$ small \\
	\noalign{\smallskip}\hline\noalign{\smallskip}
	$m_{H_3} <$ 400 & $H_1 H_2$ (BMP1) & $b \bar{b}$ (BMP3) \\
	\noalign{\smallskip}\hline\noalign{\smallskip}
 	$m_{H_3} >$ 400 & $t \bar{t}$ (BMP2) & $b \bar{b}$ + $\chi_1^\pm, \chi_2^\pm$ (BMP4)  \\
	\noalign{\smallskip}\hline
\end{tabular}
\end{table}
These features have been summarized in Table \ref{t1}. One additional feature of \textit{scenario II} is the possibility to decay into gauginos, which is related to the value of $\mu_{eff}$. This value is fixed in the CMSSM by EWSB and is usually large compared to $M_1$, leading to  the lightest neutralinos and charginos to be gaugino-like.
In the NMSSM  $\mu_{eff}$ is related to the vev of the Higgs singlet and is a free parameter. 
As mentioned above, the fit within the 3D Higgs mass parameter space is not unique. To make sure that the fit is not locked in a local instead of a global minimum we also put a grid in the 6D parameter space and fitted for each bin in the $\lambda-\kappa$ plane the remaining parameter $\tan\beta,A_\lambda,A_\kappa,A_0$ and $\mu_{eff}$. We checked that the range of resulting branching ratios  is compatible with the results from the 3D Higgs mass scan, where all parameters were left free simultaneously.
\begin{figure}
\begin{center}
\includegraphics[width=0.49\textwidth]{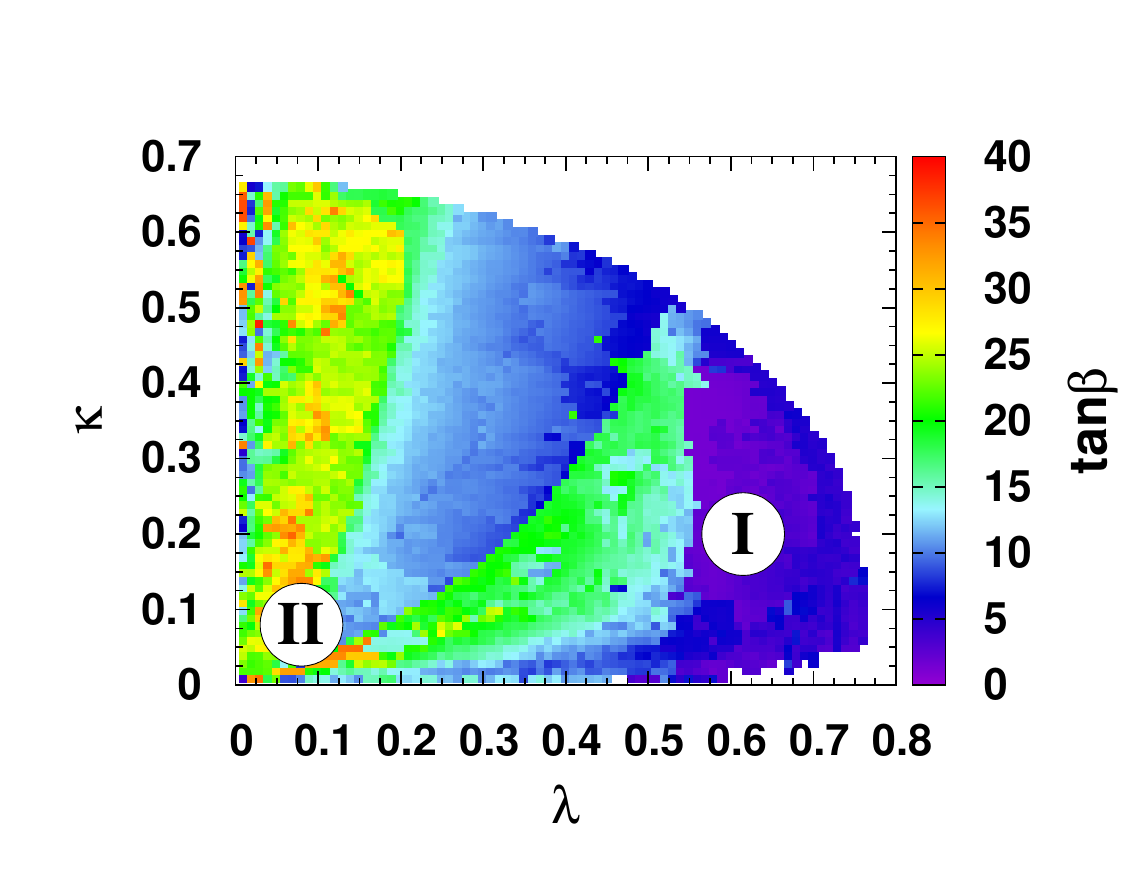}
\includegraphics[width=0.49\textwidth]{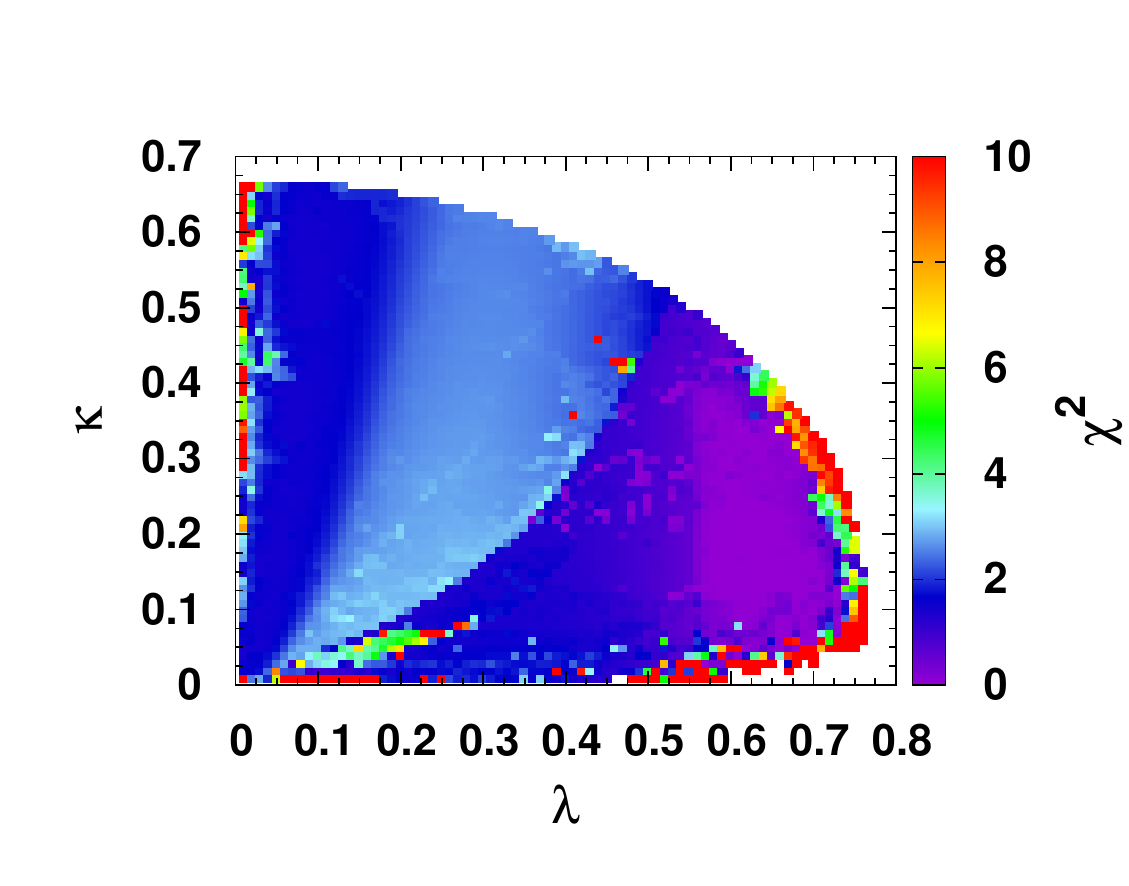}
\caption[]{$\lambda-\kappa$ plane for a fixed mass point $m_0=m_{1/2}=1000$ GeV. The shaded (color) coding corresponds to the best fit value of $\tan\beta$ (left) and the absolute value of $\chi^2$ (Eq. \ref{eq5}) (right). 
Either the lightest or the second-lightest Higgs boson is allowed to correspond to the observed Higgs boson. 
}
\label{f1}
\end{center}
\end{figure}
\begin{table}[!t]
\small
\centering
\caption{ NMSSMTools 4.6.0 input parameters for  BMP 1-4. 
The SM input parameters are $m_t(pole)=173.07$ GeV, $m_b(m_b)=4.18$ GeV and $\alpha_s(M_Z)=0.1185$. For the precision of the Higgs masses the option 8 2 was used in NMSSMTools, which means that the full one loop and the full two loop corrections from top and bottom Yukawa couplings have been used. 
  \label{t2}}
\begin{tabular}{l|c|c||c|c|}
	\hline\noalign{\smallskip}
	  & BMP1 & BMP2 & BMP3 & BMP4 \\
	\noalign{\smallskip}\cline{2-5}\noalign{\smallskip}
& \multicolumn{4}{c|}{Input at the GUT scale } \\
	\noalign{\smallskip}\hline\noalign{\smallskip}
	$m_0$ in GeV & 1000.00 & 1000.00& 1000.00&  2000.00\\
 	$m_{1/2}$ in GeV & 1000.00&  1000.00& 1000.00& 600.00\\
	$A_0$ in GeV & 2666.23& 2689.82 & -2552.64& -3322.46\\
	$A_\kappa$ in GeV & 2999.60&  2888.40& -300.14& -300.06\\
	$A_\lambda$ in GeV & 2888.27& 3041.86 & -1028.98& -640.89\\
	\noalign{\smallskip}\hline\noalign{\smallskip}
& \multicolumn{4}{c|}{Input at the SUSY scale} \\
	\noalign{\smallskip}\hline\noalign{\smallskip}
	 $\lambda \cdot 10^{2}$  & 63.06& 63.97 & 0.97& 1.66\\
	$\kappa \cdot 10^{2}$  & 38.22& 32.24 & 0.93& 1.55\\
 	$\mu_{eff}$ in GeV & 156.71& 185.68 & 104.09& 106.78\\
	\noalign{\smallskip}\hline\noalign{\smallskip}
& \multicolumn{4}{c|}{Input at the EW scale} \\
	\noalign{\smallskip}\hline\noalign{\smallskip}
	$\tan\beta$ & 2.07& 2.25 & 28.79& 14.38\\
	\noalign{\smallskip}\hline\noalign{\smallskip}
& \multicolumn{4}{c|}{Output of selected masses} \\
	\noalign{\smallskip}\hline\noalign{\smallskip}
	$\tilde{t}_1$ in GeV& 1199.55 &  1265.41 & 885.33 & 582.08\\
	$\tilde{t}_2$ in GeV&   1794.28  & 1817.64 & 1599.46& 1631.86\\
	$\tilde{\chi}_1^\pm$ in GeV& 151.95 &  181.39 & 104.90 & 104.29\\
	$\tilde{\chi}_2^\pm$ in GeV& 816.18 & 816.03  & 824.31 & 514.85 \\
	$\tilde{\chi}_1^0$ in GeV& 131.47 &  150.90 & 98.90 & 94.25 \\
	$\tilde{\chi}_2^0$ in GeV& 189.23 &  217.33 & 111.46 & 115.86\\
	\noalign{\smallskip}\hline
\end{tabular}
\end{table}
\begin{table}[!t]
\small
\centering
\caption{ Masses of the Higgs bosons for  BMP 1-4 and their corresponding Higgs production cross section at 14 TeV for the dominant gluon-gluon fusion process in \textit{scenario I} and the vector boson fusion $b\bar{b}H$ for \textit{scenario II} for the neutral Higgs bosons. The production cross section for the charged Higgs is given for the bottom-gluon fusion process. Note that for gluon fusion the $A_2$ production cross section is three times larger than the $H_3$ production cross section, although the masses are  similar.   
\label{t3}}
\begin{tabular}{l|c|c||c|c|}
	\hline\noalign{\smallskip}
 & BMP1 & BMP2 & BMP3 & BMP4 \\
	\noalign{\smallskip}\cline{2-5}\noalign{\smallskip}
	 &\multicolumn{4}{c|}{Higgs masses in GeV}\\
	\noalign{\smallskip}\hline\noalign{\smallskip}
	$H_1$ & 100.0 & 100.0 & 100.0 & 100.0\\
 	$H_2$ & 125.2 & 125.2 & 123.3& 123.0 \\
	$H_3$ & 350.0 & 450.0 & 850.0& 1000.0 \\
	$A_1$ & 300.0 & 300.0 & 300.0& 300.0 \\
	$A_2$ & 341.7 & 444.9  & 850.0 &1000.0 \\
	$H^\pm$ & 334.4 & 437.3 & 854.1 & 1003.3   \\	
	\noalign{\smallskip}\hline\noalign{\smallskip}
	& \multicolumn{4}{c|}{ $\sigma_{prod,ggh} $  in pb}  \\
	\noalign{\smallskip}\hline\noalign{\smallskip}
	$H_1$ & 0.55 &  0.42 & 0.18& 0.20\\
 	$H_2$ &  46.05 &  46.3 & 46.36& 46.23\\
	$H_3$ &  2.77 &  1.44 & $<0.01$& $<0.01$\\
	$A_1$ &  0.06 &  0.10 & $<0.01$ & $<0.01$\\
	$A_2$ &  11.14 &  3.38 & $<0.01$& $<0.01$\\
	\noalign{\smallskip}\hline\noalign{\smallskip}
	& \multicolumn{4}{c|}{$\sigma_{prod,b\bar{b}h} $  in pb }  \\
	\noalign{\smallskip}\hline\noalign{\smallskip}
	$H_1$ & 0.35 &  0.25 & $<0.01$ & $<0.01$\\
 	$H_2$ & 0.60  & 0.60  & 0.66& 0.65\\
	$H_3$ & 0.06  &  0.02 & 0.21& 0.02\\
	$A_1$ & $<0.01$  & $<0.01$  &$<0.01$ & $<0.01$\\
	$A_2$ & 0.07  &  0.03 & 0.21& 0.02\\
	\noalign{\smallskip}\hline\noalign{\smallskip}
	& \multicolumn{4}{c|}{$\sigma_{prod,gb} $  in pb }  \\
	\noalign{\smallskip}\hline\noalign{\smallskip}
	$H^\pm$ &  0.37 & 0.15  & 0.01 &  $<0.01$\\
	\noalign{\smallskip}\hline
\end{tabular}
\end{table}
The transition between \textit{scenario I} and \textit{II} can be readily observed, if one plots the best fit value of $\tan\beta$ in the $\lambda-\kappa$ plane, as shown in the top left panel of Fig. \ref{f1}. The dark (blue) regions for $\lambda \geq 0.55$ corresponds to \textit{scenario I}, while the shaded (greenish) regions for  $\lambda \leq 0.1$ corresponds to \textit{scenario II}. The right panel of Fig. \ref{f1} shows the $\chi^2$ function of Eq. \ref{eq5} without the $\chi^2_{H_3}$ term, since $m_{H_3}$ was allowed to vary in the plane. The region between the two greenish regions has a poorer $\chi^2$ value, which originates from the fact, that neither the lightest nor the second lightest NMSSM Higgs boson has the right mass and right couplings in comparison with the observed Higgs boson. The white region within the $\lambda-\kappa$ plane is not allowed, since for such large values of the parameters one reaches a Landau pole. For the benchmark points we choose a typical point in regions I and II (indicated by I and II  in the left panel of Fig. \ref{f1}). The corresponding parameter set and sparticle masses are given in Table \ref{t2}. These benchmark points are each characterized by a specific branching ratio being dominant, as will be discussed later. The Higgs boson masses and LHC production cross sections for the four benchmark points have been summarized in Table \ref{t3}.

\begin{figure}
\begin{center}
\includegraphics[width=0.48\textwidth]{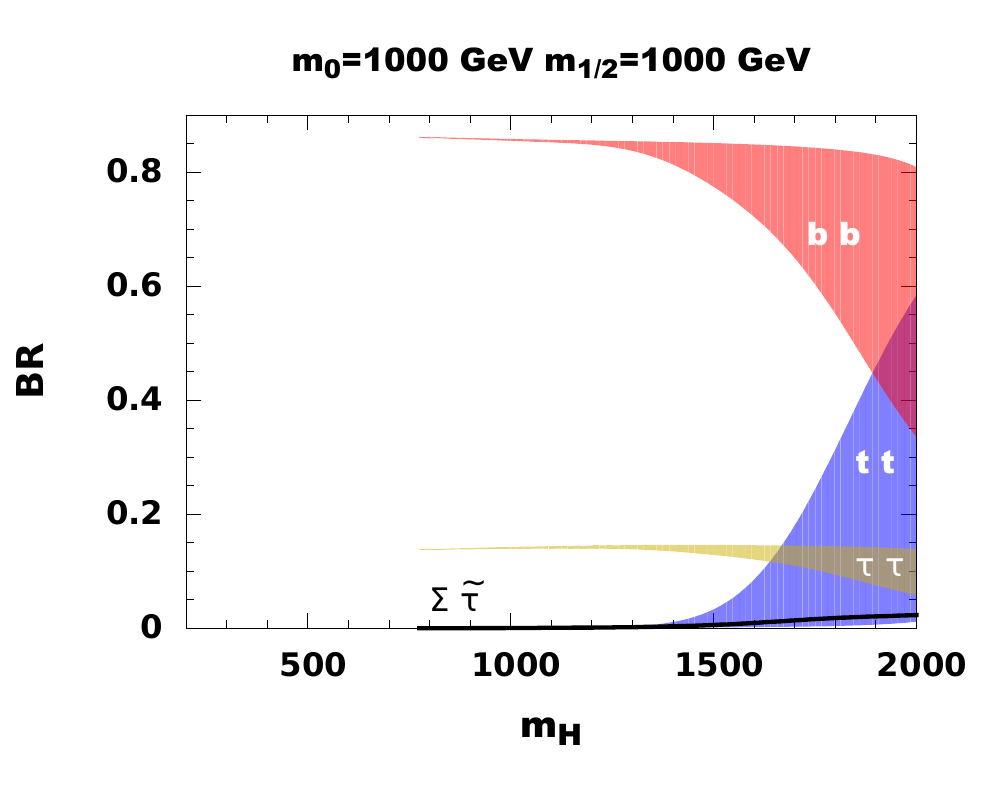}
\includegraphics[width=0.48\textwidth]{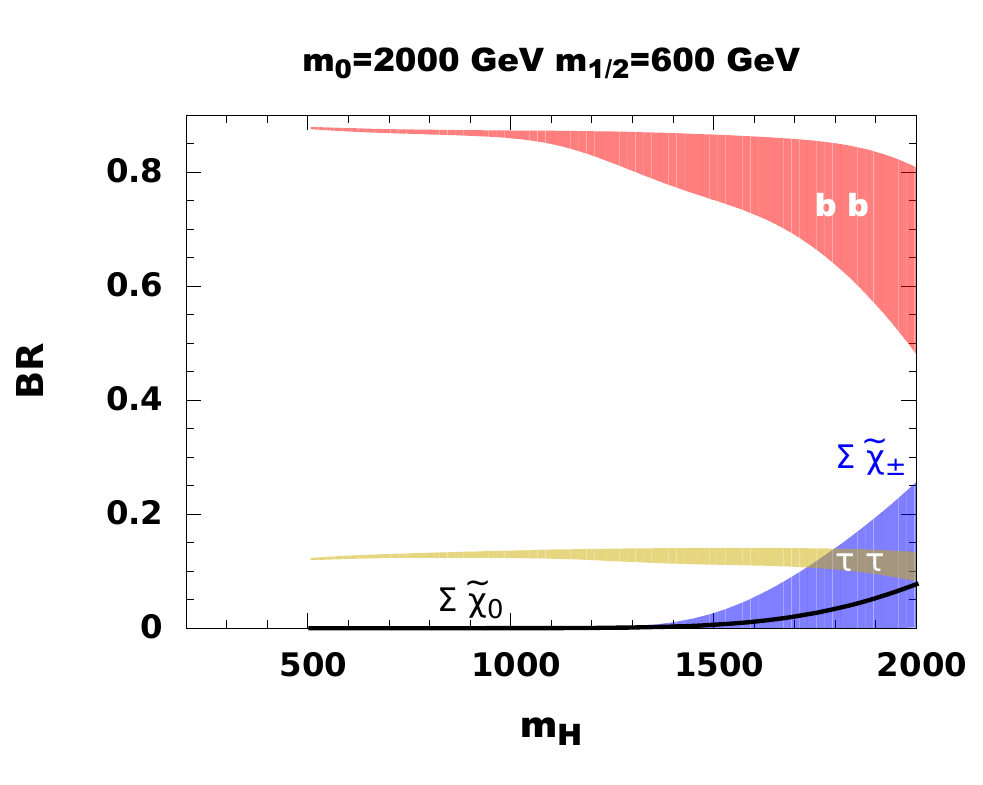}
\caption[]{ The  branching ratios of the heavy Higgs boson $H$ in the CMSSM as function of its mass. The branching ratios for the pseudoscalar Higgs boson are similar. The dominant branching ratios are shown here as bands, while the width of the bands include the variation of $A_0$ and $\tan\beta$. For clarity the branching ratios into staus and neutralinos are shown as lines without band which represent the mean of the corresponding band. The decay into $t\overline{t}$ is suppressed by the large value of $\tan\beta$, required by the relic density in most of the parameter space. 
For a lower choice of $m_{1/2}$ the branching ratios into charginos and neutralinos open up, as shown in the right-hand panel. 
}
\label{f2}
\end{center}
\end{figure}

\subsection{LHC limits on Higgs boson masses}
Apart from the observation of the SM-like Higgs boson at 125 GeV the LHC has not observed any other Higgs bosons, but placed limits on the heavy Higgs bosons. In SUSY the production cross section for the heavy Higgs boson is proportional to $\tan^2\beta$ (see e.g. \cite{Beskidt:2010va}), so the limits are a strong function of $\tan\beta$ \cite{Mitani:2015aga,Khachatryan:2014wca}. Typically, heavy pseudo-scalar Higgs boson below 800 GeV are excluded for $\tan\beta \geq 45$, but no limits are obtained for $\tan\beta \le 4$.
 Furthermore, the constraints from B-physics have to be taken into account.  The $b_s\rightarrow \mu\mu$ decay modes (proportional to $\tan^6\beta$) requires rather heavy SUSY masses for large  $\tan\beta$ or, alternatively, a small mass splitting in the stop sector, see e.g. \cite{Beskidt:2011qf}. Not only $b_s\rightarrow \mu\mu$ but also $b \rightarrow s \gamma$, restricts the allowed parameter space, so to be in agreement with the B-physics constraints we chose $\tan\beta$ to be not larger than 30 for our benchmark points.  
The absolute lower limits of the heavier Higgs masses are given by the Higgs boson of 125 GeV.  An additional lower limit on the heavier Higgs boson mass around 800 GeV exists in both scenarios. In \textit{scenario I} this limit results from the relic density constraint if the correct relic density is required. Below this limit the relic density is too small, which is allowed if  dark matter has contributions from particles different from the LSP.  In \textit{scenario II} (large $\tan\beta$) the limit comes from the LHC, as discussed above.

\subsection{Heavy Higgs branching ratios within the CMSSM}
Before discussing the branching ratios in the NMSSM, we discuss the simpler case of the CMSSM, where only two free parameters ($A_0$ and $\tan\beta$) enter the Higgs sector.
The branching ratios of the heavy Higgs bosons were calculated with SUSY-HIT \cite{Djouadi:2006bz} for a grid in the $A_0-\tan\beta$ plane and are plotted in Fig. \ref{f2} 
for two CMSSM mass points not excluded by the LHC ($m_0$=1000/2000 GeV, $m_{1/2}$=1000/600 GeV left/right-hand side). 
The last mass point corresponds to a lower value of  $m_{1/2}$, which leads to lower gaugino masses.
The branching ratios to gauginos become  important for high values of the Higgs mass, as shown in the right panel of Fig. \ref{f2}.  For higher values of  $m_{1/2}$ the branching ratio into top quark pairs becomes dominant at large Higgs boson values, as shown in the left panel.
For $m_H<1.5$ TeV the branching ratios into b-quarks and tau-leptons always dominate.
This is easily understood as follows: at tree level the heavy pseudo-scalar Higgs boson mass is given by the sum of the mass terms in the Higgs potential, i.e. $m_1^2+m_2^2$. The $m_2^2$ parameter is driven negative by the large corrections from the top Yukawa coupling $h_t$ and induces EWSB. However, $m_1^2$ gets also large negative  corrections from the bottom Yukawa coupling $h_b$, which can become comparable to  $h_t=m_t/v_2$ for  large values of $\tan\beta$, since $h_b= m_b/v_1=m_b\tan\beta/ v_2$. Hence, for large values of $\tan\beta$ $m_1^2$ and $m_2^2$ both become small by negative corrections of $h_b$ and $h_t$, respectively, thus leading to small values of $m_A$ and enhancing at the same time the branching into down-type fermions. So the heaviest Higgs bosons are expected to decay into b-quarks and $\tau$-leptons for masses below 1.5 TeV, which is close to the reach at the LHC  \cite{Aad:2011rv}. Masses above 1.5 TeV require smaller values of $\tan\beta$ in order to increase $m_2^2$. These smaller $\tan\beta$ values allow branchings into other channels. The widths of the bands originate mainly from the allowed variation of $A_0$ and $\tan\beta$ for a given mass. 

\begin{figure}
\begin{center}
\includegraphics[width=0.44\textwidth]{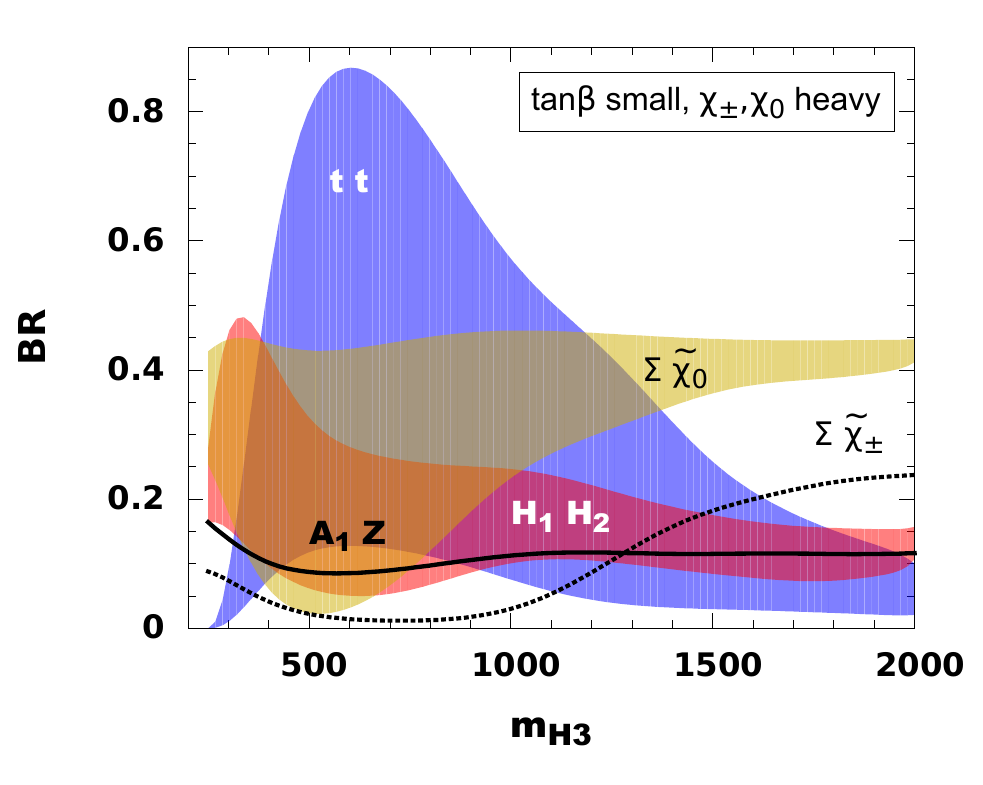}
\includegraphics[width=0.44\textwidth]{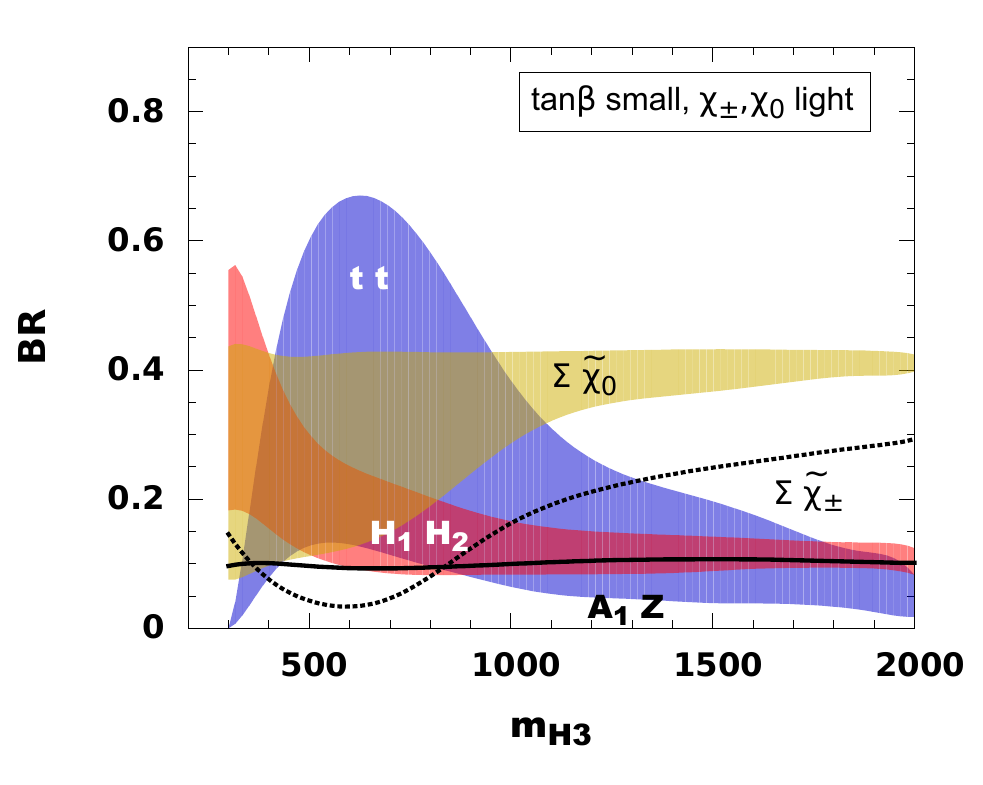}
\includegraphics[width=0.44\textwidth]{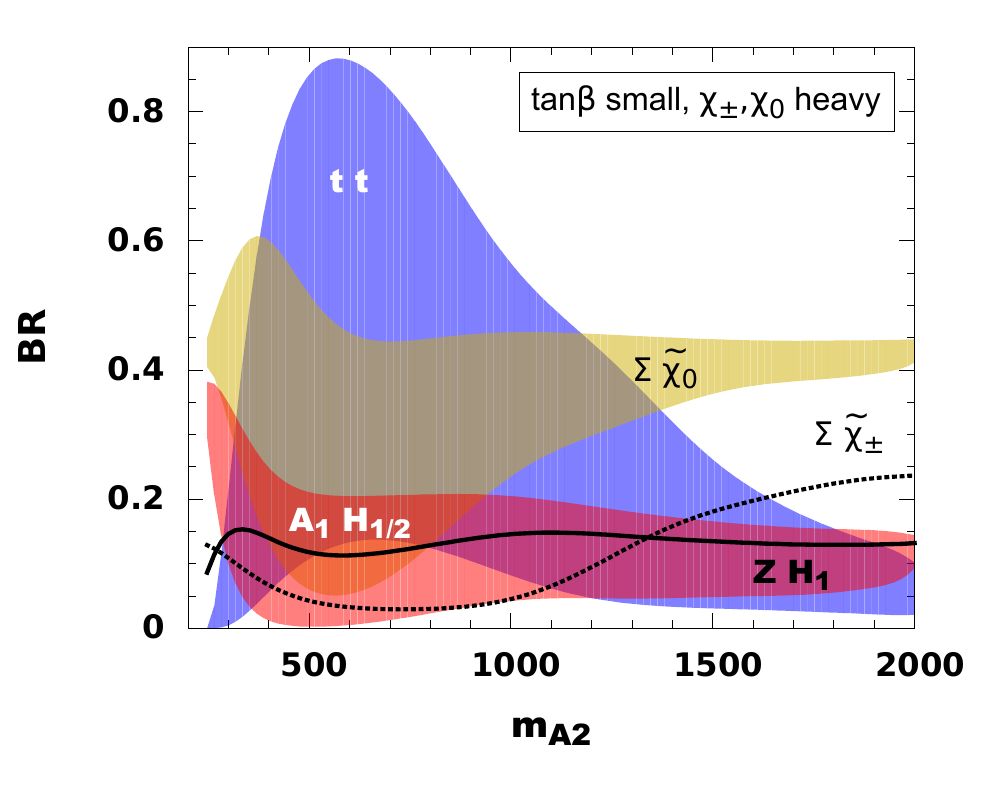}
\includegraphics[width=0.44\textwidth]{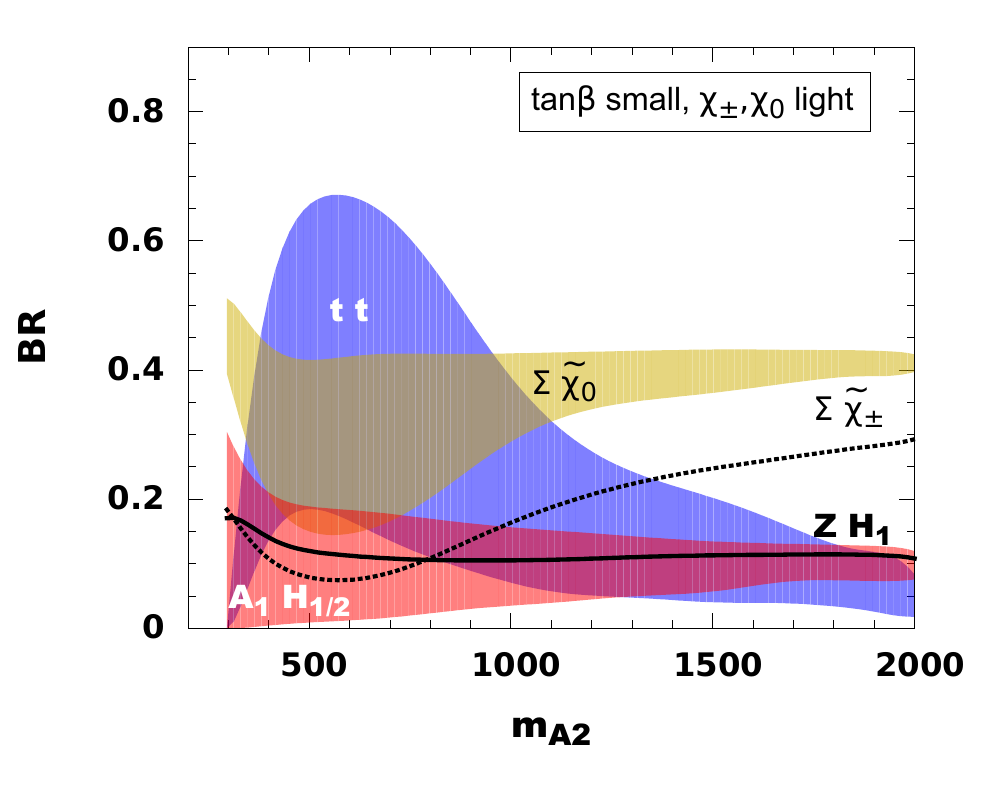}
\includegraphics[width=0.44\textwidth]{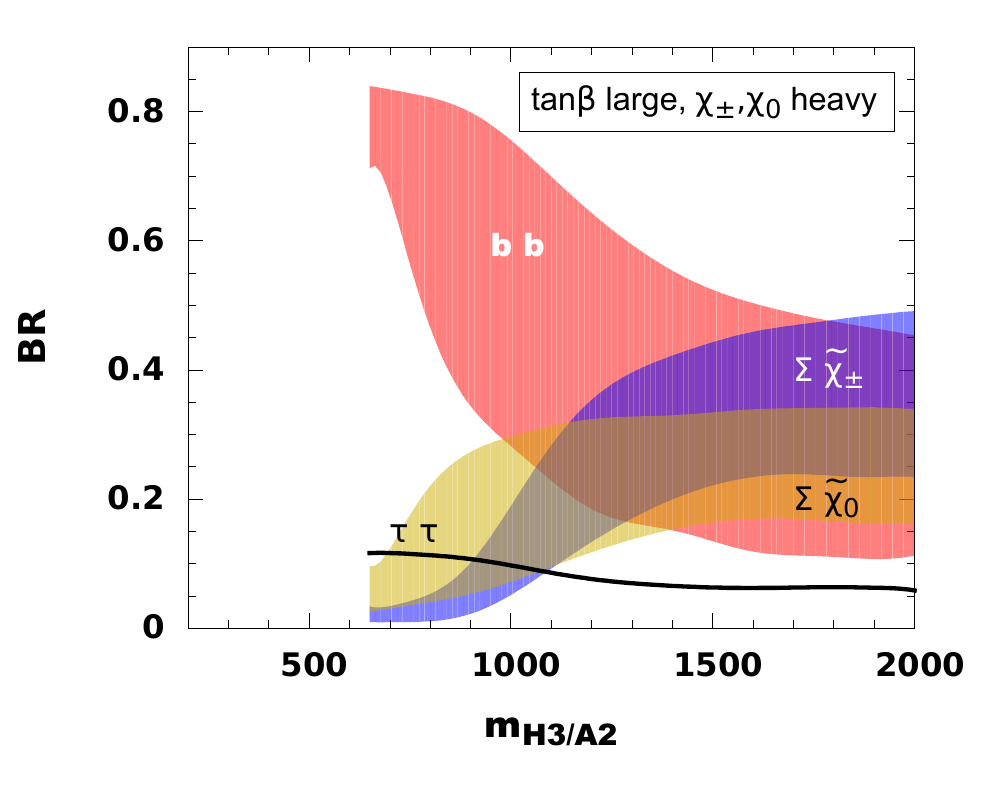}
\includegraphics[width=0.44\textwidth]{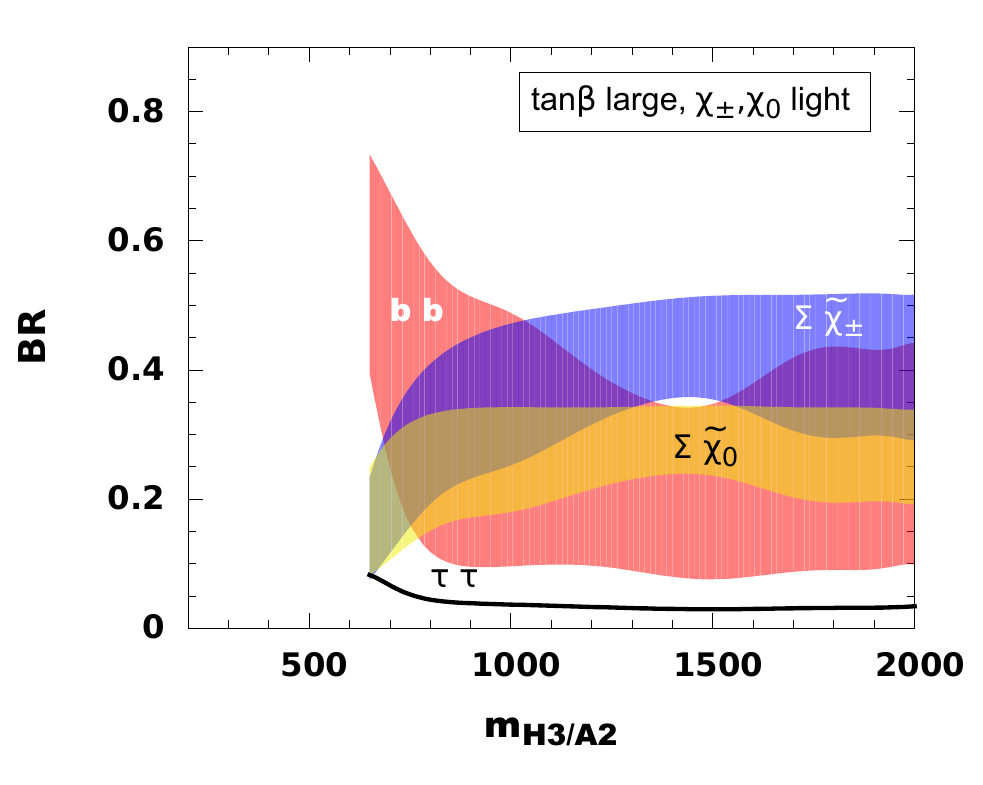}
\small 
\caption[]{ The  branching ratios of a heavy Higgs boson in the NMSSM as function of its mass for \textit{scenario I} (top,middle) and \textit{II} (bottom).  For \textit{scenario II} the branching ratios for $H_3$ and $A_2$ are similar, so they have been plotted together in the last row. 
The main difference between the branching ratios of $H_3$ and $A_2$ in \textit{scenario I} are the additional decays of $A_2$  into $A_1 H_{1/2}$ (orange band) and $Z H_1$ (solid black line) . These decays are not allowed for the scalar Higgs boson $H_3$.
The dominant branching ratios are shown  as bands, where the width of the bands represents the allowed variation of the  NMSSM parameters. To simplify the plot the smaller branching ratios have  been shown as  a line representing the average of the band. The decays into gauge boson pairs is negligible in both scenarios, while $b\overline{b}$ and $\tau\overline{\tau}$ are important in \textit{scenario II} with large $\tan\beta$. Decays into gaugino masses become possible as well, if they are light enough. Here they were chosen to correspond to CMSSM mass points not excluded by the LHC ($m_0$=1000/2000,  $m_{1/2}$=1000/600 GeV left/right-hand side).  }
\label{f3}
\end{center}
\end{figure}

\begin{table}[!t]
\small
\centering
\caption{Summary of heavy Higgs boson branching ratios  (in \%) for BMP 1-4. \label{t4}}
\begin{tabular}{l|c|c||l|c|c}
	\hline\noalign{\smallskip}
	  & BMP1 & BMP2 & & BMP3 & BMP4 \\
	\noalign{\smallskip}\hline\noalign{\smallskip}
	$H_3 \rightarrow H_1 H_2$ &  67.8 & 25.4&$H_3 \rightarrow \tau \tau$ & 12.2 & 3.7\\
	$H_3 \rightarrow t \bar{t}$ & 6.8 &  54.8&$H_3 \rightarrow t \bar{t}$ &  0.4 & 1.9\\ 
	$H_3 \rightarrow H_1 H_1$ & 6.1 & 1.8&$H_3 \rightarrow b \bar{b}$ &  81.4&  24.5\\
	$H_3 \rightarrow A_1 Z$ & - & 1.3&$H_3 \rightarrow \chi_1^0 \chi_1^0$ & 0.4& 1.8\\
	$H_3 \rightarrow \chi_1^0 \chi_1^0$ & 7.4 & 8.1&$H_3 \rightarrow \chi_1^0 \chi_4^0$ & 1.6& 2.7\\
	$H_3 \rightarrow \chi_1^0 \chi_2^0$ & 0.4& 0.7&$H_3 \rightarrow \chi_1^0 \chi_5^0$ & -&7.6\\
	$H_3 \rightarrow \chi_1^0 \chi_3^0$ & - & 4.2& $H_3 \rightarrow \chi_2^0 \chi_2^0$ & - &0.6 \\
	$H_3 \rightarrow \chi_1^+ \chi_1^-$ & 5.3& 1.9& $H_3 \rightarrow \chi_2^0 \chi_4^0$ & 2.5& 3.9\\
	  \multicolumn{3}{c||}{}&	$H_3 \rightarrow \chi_2^0 \chi_5^0$ & - &11.8\\
	  \multicolumn{3}{c||}{}&	$H_3 \rightarrow \chi_1^+ \chi_1^-$ & 1.1& 3.6\\
	  \multicolumn{3}{c||}{}&		$H_3 \rightarrow \chi_1^+ \chi_2^-$ & -&18.6\\
	  \multicolumn{3}{c||}{}&		$H_3 \rightarrow \chi_2^+ \chi_1^-$ & -&18.6\\
	\noalign{\smallskip}\hline\noalign{\smallskip}
	$A_2 \rightarrow A_1 H_2$ & - & 1.2&$A_2 \rightarrow \tau \tau$ & 12.2&  3.8\\
	$A_2 \rightarrow t \bar{t}$ &  - & 63.9&$A_2 \rightarrow t \bar{t}$ &  0.4& 2.0\\ 
	$A_2 \rightarrow A_1 H_1$ & - & 1.1&$A_2 \rightarrow b \bar{b}$ &  81.3& 24.6\\
	$A_2 \rightarrow H_1 Z$ & 49.4& 12.9&$A_2 \rightarrow \chi_1^0 \chi_1^0$ & 0.5\\
	$A_2 \rightarrow \chi_1^0 \chi_1^0$ & 35.7& 12.1&$A_2 \rightarrow \chi_1^0 \chi_4^0$ & 2.6& 2.4\\
	$A_2 \rightarrow \chi_1^0 \chi_2^0$ & 0.4 & 0.8 & $A_2 \rightarrow \chi_1^0 \chi_4^0$ & - &3.7 \\
	$A_2 \rightarrow \chi_1^0 \chi_3^0$ & -  & 1.6 &$A_2 \rightarrow \chi_1^0 \chi_5^0$ & -& 12.7 \\
	$A_2 \rightarrow \chi_1^+ \chi_1^-$ & 10.2& 3.3& $A_2 \rightarrow \chi_2^0 \chi_2^0$ & -& 0.5\\
	  \multicolumn{3}{c||}{} &$A_2 \rightarrow \chi_2^0 \chi_4^0$ & 1.5 & 2.6\\
	  \multicolumn{3}{c||}{} & $A_2 \rightarrow \chi_2^0 \chi_5^0$ & - &6.4\\
	  \multicolumn{3}{c||}{}& 	$A_2 \rightarrow \chi_4^0 \chi_4^0$ & - &0.1\\
	  \multicolumn{3}{c||}{}& 	$A_2 \rightarrow \chi_4^0 \chi_5^0$ & - & 0.3 \\
	  \multicolumn{3}{c||}{}& $A_2 \rightarrow \chi_1^+ \chi_1^-$ & 1.2& 4.1\\
	  \multicolumn{3}{c||}{} & 	$A_2 \rightarrow \chi_1^+ \chi_2^-$ & - &18.3\\
	  \multicolumn{3}{c||}{}& 	$A_2 \rightarrow \chi_2^+ \chi_1^-$ & -& 18.3\\
	\noalign{\smallskip}\hline\noalign{\smallskip}
	$H^\pm \rightarrow t \bar{b}$ & 83.3 & 73.7 &$H^\pm \rightarrow \tau \nu_\tau$ & 12.4 &  3.8\\
	$H^\pm \rightarrow W^\pm H_1$ &  13.9& 15.8 & $H^\pm \rightarrow t \bar{b}$ & 82.4 & 26.6\\
	$H^\pm \rightarrow \chi_1^\pm \chi_1^0$ & 2.7 &  6.7 & $H^\pm \rightarrow \chi^\pm_1 \chi^0_4$ & 4.9 &10.5 \\
	$H^\pm \rightarrow \chi_1^\pm \chi_3^0$ & - & 2.1 & $H^\pm \rightarrow \chi_1^\pm \chi_5^0$ & - &  18.8 \\
	  \multicolumn{3}{c||}{}& 	$H^\pm \rightarrow \chi_2^\pm \chi_1^0$ & -& 21.3\\
	  \multicolumn{3}{c||}{}& 	$H^\pm \rightarrow \chi_2^\pm \chi_1^0$ & -& 18.0   \\
	\noalign{\smallskip}\hline
\end{tabular}
\end{table}

\begin{table}[!t]
\small
\centering
\caption{As in Table \ref{t4}, but for the branching ratios (in \%) of the lighter Higgs bosons. \label{t5}}
\begin{tabular}{l|c|c||l|c|c}
	\hline\noalign{\smallskip}
	  & BMP1 & BMP2 & & BMP3 & BMP4 \\
	\noalign{\smallskip}\hline\noalign{\smallskip}
	$H_1 \rightarrow b \bar{b}$ &  90.1& 90.1&$H_1 \rightarrow b \bar{b}$ &  70.6&  60.9\\
	$H_1 \rightarrow \tau \tau$ & 9.5& 9.5&$H_1 \rightarrow c \bar{c}$ &  8.3&  12.1\\
	  \multicolumn{3}{c||}{} & $H_1 \rightarrow \tau \tau$ & 7.2 & 6.2\\
	  \multicolumn{3}{c||}{} & $H_1 \rightarrow g g$ & 11.9 & 17.7\\
	  \multicolumn{3}{c||}{}  & $H_1 \rightarrow W W$ & 1.5  & 2.2\\
	\noalign{\smallskip}\hline\noalign{\smallskip}
	$H_2 \rightarrow b \bar{b}$ &  62.4& 62.1 & &65.9&  66.0\\
	$H_2 \rightarrow W W$ &  19.8 & 20.0& &16.6&  16.4\\
	$H_2 \rightarrow g g$ & 5.6& 5.7& &5.4& 5.5\\
	$H_2 \rightarrow \tau \tau$ & 6.7& 6.7& &7.1& 7.1\\
	$H_2 \rightarrow c \bar{c}$ &  2.8&  2.8& & 2.8& 2.9\\
	$H_2 \rightarrow Z Z$ & 2.2 &  2.2& & 1.7& 1.7 \\
	\noalign{\smallskip}\hline\noalign{\smallskip}
	$A_1 \rightarrow g g$ &  0.0& 1.2&$A_1 \rightarrow \chi_1^0 \chi_1^0$ & 26.1& 25.7\\
	$A_1 \rightarrow \tau \tau$ &  0.0&  1.8&$A_1 \rightarrow \chi_2^0 \chi_2^0$ & 23.6& 23.3\\
	$A_1 \rightarrow b \bar{b} $ & 0.0 &  14.3 &$A_1 \rightarrow \chi_1^+ \chi_1^-$ & 50.2& 50.7\\
	$A_1 \rightarrow Z H_1$ & 0.4 & 82.9& 	  \multicolumn{3}{}{}  \\
	$A_1 \rightarrow \chi_1^0 \chi_1^0$ & 99.6 & -& \multicolumn{3}{}{} \\
	\hline\noalign{\smallskip}
\end{tabular}
\end{table}

\subsection{Heavy Higgs branching ratios in the NMSSM}

The large difference in the branching ratios of the heavy Higgs boson between the NMSSM and CMSSM is clear from a comparison of   Figs. \ref{f2} and \ref{f3}. The latter   shows  the branching ratios of the heavy scalar and pseudo-scalar Higgs bosons as function of their masses in the NMSSM, again for the two CMSSM mass points discussed before.

 In the CMSSM  the scalar and pseudo-scalar heavy bosons have similar branching ratios, but in the NMSSM one has two  scalar Higgs bosons with a mass below the heaviest one, so the heaviest one may decay into the two lighter ones ($H_3\rightarrow H_1 H_2$), if kinematically allowed. This is  forbidden by parity conservation for the pseudo-scalar boson. Therefore, $H_3$ and $A_2$ have different branching ratios, as can be seen from Fig. \ref{f3}.  In the NMSSM the Higgs boson masses are largely independent of $\tan\beta$, so  for each mass considered both  scenarios   are possible, as shown in the different rows. 
 The width of the bands corresponds mainly to the allowed variation of $\lambda$ and $\kappa$.  The variation of the lightest pseudo-scalar Higgs boson mass $m_{A_1}$ between 25 and 500 GeV gives a smaller contribution to the width of the bands.  

The bottom row with large $\tan\beta$ is similar to the branching ratios in the CMSSM (Fig. \ref{f2}), i.e. large branching ratios into down-type fermions. They differ  because of the chosen small values of $\mu_{eff}$ in the NMSSM, which leads to lighter neutralinos and charginos  in comparison with the CMSSM, where $\mu$ is large due to EWSB.  The lightest  charginos and neutralinos in the NMSSM are in addition Higgsino and singlino-like in contrast to the bino and wino-like sparticles in the CMSSM. The threshold for the gauginos depends on $m_{1/2}$, as can be seen from a comparison of the left and right panels in  Fig. \ref{f3}. Only the sum of the branching ratios into either charginos or neutralinos has been indicated.

 For low values of $\tan\beta$  the decay modes into b-quarks and tau-leptons  are typically absent and the decays into top quarks (when above threshold) or lighter Higgs bosons prevail, as  can be seen from the top row in Fig. \ref{f3}. For the pseudo-scalar Higgs mass the decay into two lighter scalar Higgs bosons is forbidden, so the main decay modes are into top quarks and gauginos, as shown in the middle row of Fig. \ref{f3}. If $\tan\beta$ is large (\textit{scenario II}) the dominant decay are into down-type fermions and gauginos, if kinematically allowed, as shown in the bottom row of Fig. \ref{f3}.

Within the bands of the possible branching ratios we propose two benchmark points for each scenario: one in which the heavy scalar Higgs decays mostly into $H_1 H_2$ (called BMP1) and one in which $H_3$ decays mostly into $t\bar{t}$ (called BMP2) for \textit{scenario I}. In \textit{scenario II} BMP3 corresponds to a do\-minant decay into a pair of b quarks. In BMP4 the decay into $b \bar{b}$ is reduced due to the significant decay into charginos and neutralinos. 
The heavy pseudo-scalar Higgs mass is almost degenerate in mass with the heavy scalar one, so they will be produced simultaneously, but with different branching ratios and cross sections. The masses and cross sections have been summarized  before in Table \ref{t3}. 
Numerical values of the branching ratios for the benchmark points are listed in Tables \ref{t4} and \ref{t5}.
The production cross section for the neutral Higgs bosons has been calculated for 14 TeV using SusHi \cite{Harlander:2012pb,Harlander:2002wh,Harlander:2003ai,Aglietti:2004nj,Bonciani:2010ms,Degrassi:2010eu,Degrassi:2011vq,Degrassi:2012vt,Liebler:2015bka}. The cross section for the charged Higgs boson at 14 TeV has been estimated using FeynHiggs \cite{Hahn:2013ria,Frank:2006yh,Degrassi:2002fi,Heinemeyer:1998np,Heinemeyer:1998yj}. 
\textit{Scenario I} is dominated by the gluon fusion production cross section, while for \textit{scenario II} with large $\tan\beta$ the $b\bar{b}H$ cross section dominates. Since the cross sections for charged Higgs production originate from the same diagrams in the MSSM and NMSSM, the values for the MSSM, as calculated with FeynHiggs, were taken. In the following we discuss some of the features of these benchmark points. 

\subsubsection{Benchmark point BMP1 with $H_1 H_2$  decay dominant in \textit{scenario I}}
 The  $H_3$ and $A_2$ bosons have practically the same mass  (350 and 342 GeV, respectively), but they have quite different decays: $H_3$ decays for 68\% into $H_1 + H_2$, while $A_2$ decays for 49\% into $H_1 + Z$  and the remaining decay modes are largely gauginos but the production cross section of $A_2$ is 3 times larger compared to $H_3$, see Table \ref{t3}. The decay mode of the lightest pseudo-scalar Higgs boson $A_1$, shown in Table \ref{t5}, is not  $Z+H_1$, as in BMP2 (although the masses of the lighter Higgs bosons are identical), but the main decay mode is now into LSPs, so an invisible final state.  This benchmark point is characterized by a large fraction of double Higgs production in the $H_2$ decay,  while the $A_2$  decays into  $Z+H_1$ or gauginos, either neutral or charged, which in turn have a rich spectrum of decay modes.  The $A_1$ boson decays largely into invisible neutralinos, while the lightest Higgs boson $H_1$  decays largely into $b\overline{b}$ and tau-pairs. The charged Higgs boson decays largely into $t\bar{b}$ and $W^\pm H_1$.

\subsubsection{Benchmark point BMP2 with $t\overline{t}$  decay dominant in \textit{scenario I}}
The  $H_3$ and $A_2$ bosons have similar masses   (450 and 446 GeV, respectively). In both cases the $t\bar{t}$ decay is dominant, so the cross sections can be added. Note that $H_3$ can decay into $H_1+ H_2$ as well, while for $A_2$ the decay into $H_1+ Z$ and LSPs yields the second largest branching ratio. $A_1$ decays largely into $Z+H_1$,  as shown in Table \ref{t5}. So this benchmark point is characterized by a large fraction of $t \bar{t}$ final states, which can be searched for as a broad bump around 450 GeV in the tail of the $t \bar{t}$ invariant mass spectrum. Furthermore, events with two Z bosons and the $H_1$  Higgs boson of 100 GeV with practically SM decay modes can be searched for from the $A_2$ decay mentioned above.  As can be seen from Table \ref{t4}, the dominant decay mode for the charged Higgs is into $t\bar{b}$ and $W^\pm H_1$.

\subsection{Benchmark point BMP3 with $b\overline{b}$  decay dominant in \textit{scenario II}}
For this benchmark point the chosen masses of the heavy Higgs boson are heavier in comparison to BMP1 and BMP2. The branching ratios of $H_3$ and $A_2$ are shown in Table \ref{t4}. The mass splitting for such heavy Higgs boson masses is negligible. In both cases the $b\bar{b}$ decay is dominant, so the cross sections can be added. But since this channel has a large background the smaller branching ratio into $\tau$ leptons with a smaller background may be the preferred search channel  for the heavy Higgs boson.
$A_1$ decays largely into charginos and neutralinos,  as shown in Table \ref{t5}.  Although the mass of the charged Higgs boson is heavier compared to BMP1 and BMP2, the decay into  $t\bar{b}$ and $\tau \nu_{\tau}$ is dominant, because of the heavy charginos and neutralinos.

\subsection{Benchmark point BMP4 with $\chi_1^\pm\chi_2^\pm$  decay dominant in \textit{scenario II}}
The last benchmark point has heavy Higgs boson masses around 1 TeV. The mass difference for $H_3$ and $A_2$ is negligible  and their branching ratios are shown in Table \ref{t4}. The $b\bar{b}$ decay is still significant, but the decay into charginos starts to dominate. Since the decay mode of the dominating branching ratio includes $\chi_2^\pm$ one expects gauge bosons from its decay. 
Invisible decays are expected from $A_1$, which decays largely into charginos and neutralinos,  as shown in Table \ref{t5}. For the charged Higgs boson the decay into charginos and missing transverse energy from the neutralinos starts to dominate, so the decay into $t\bar{b}$ decreases in comparison with the other benchmark points.

\section{Conclusion}
We surveyed the branching ratios of the Higgs bosons in the constrained minimal and next-to minimal supersymmetry scenarios. To limit the para\-meter space we restricted ourselves to the well-motivated common GUT scale masses for the SUSY partners, but the Higgs boson masses and their branching ratios are largely independent of the GUT scale constraints.
The interest in the next-to-minimal scenario with an additional singlet stems among others from the  increase at tree level of the SM-like Higgs boson, so the 125 GeV does not need large radiative corrections from stop loops. In addition, the $\mu$-parameter in the NMSSM is naturally of the order of the electroweak scale, thus avoiding the $\mu$-problem \cite{Ellwanger:2009dp}. However, the Higgs sector has now 6 free parameters. This 6D parameter space  makes it difficult to obtain insight in the possible range of masses and branching ratios. To solve this problem we considered instead the parameter space of the 6 Higgs masses, which reduces to a 3D mass space, if one takes into account that one Higgs mass has to be 125 GeV and the heavy Higgs bosons are practically mass-degenerate. By projecting the 6D parameter space of the NMSSM Higgs sector on the 3D parameter space of the masses we obtained the range of branching ratios of each Higgs boson mass  in two typical scenarios, as shown in Table \ref{t1}. Two benchmark points for each scenario have been presented, which can be used to search for signatures distinguishing the MSSM and NMSSM.

The recent diphoton excess by CMS \cite{CMS-PAS-EXO-15-004} and ATLAS \cite{ATLAS-CONF-2015-081} may hint for a new particle with a mass around 750 GeV, which is in agreement with the allowed mass range for the heavy Higgs bosons. Due to the large mass many decay channels are possible, so the loop induced decay into photons leads to a branching ratio of the order of $10^{-5}$. The number of expected events is then well below one. However, about 10 have been observed in {\it both} experiments at a similar mass, which makes it difficult to  dismiss the excess as a statistical fluctuation. The large discrepancy with the expected NMSSM cross section makes it also difficult to interpret the excess in the framework of SUSY, but many other explanations have been proposed, see e.g. \cite{Ellis:2015oso,King:2016wep,Petersson:2015mkr,Bhattacharya:2016lyg,Chang:2015sdy}. Fortunately, future data will soon reveal if these are fluctuations or new physics.

\section*{Acknowledgements}
Support from the Heisenberg-Landau program and the Deutsche Forschungsgemeinschaft (DFG, Grant BO 1604/3-1)  is warmly acknowledged. We thank  S. Heinemeyer for information on charged Higgs production in FeynHiggs. We thank the anonymous referee for the suggestions to improve the manuscript.


\bibliographystyle{lucas_unsrt}
\bibliography{mybib}

\providecommand{\href}[2]{#2}\begingroup\raggedright\begin{thebibliography}{10}%
\makeatletter
\providecommand{\hrefCMSnoop }[0]{\@secondoftwo}%
\makeatother

\bibitem{Haber:1984rc}
\hrefCMSnoop {} {H.~E. Haber and G.~L. Kane, ``{The Search for Supersymmetry:
  Probing Physics Beyond the Standard Model}'',} \textit{ Phys.Rept.} \textbf{
  117} (1985)
75--263.

\bibitem{deBoer:1994dg}
\hrefCMSnoop {} {W.~de~Boer, ``{Grand unified theories and supersymmetry in
  particle physics and cosmology}'',} \textit{ Prog.Part.Nucl.Phys.} \textbf{
  33} (1994) 201--302,
\href{http://www.arXiv.org/abs/hep-ph/9402266}{\texttt{ arXiv:hep-ph/9402266}}.

\bibitem{Martin:1997ns}
\hrefCMSnoop {} {S.~P. Martin, ``{A Supersymmetry primer}'',} \textit{
  Perspectives on supersymmetry II, Ed. G. Kane} (1997)
\href{http://www.arXiv.org/abs/hep-ph/9709356}{\texttt{ arXiv:hep-ph/9709356}}.

\bibitem{Aad:2012tfa}
\hrefCMSnoop {} {{ ATLAS} Collaboration, ``{Observation of a new particle in
  the search for the Standard Model Higgs boson with the ATLAS detector at the
  LHC}'',} \textit{ Phys.Lett.} \textbf{ B716} (2012) 1--29,
\href{http://www.arXiv.org/abs/1207.7214}{\texttt{ arXiv:1207.7214}}.

\bibitem{Chatrchyan:2012ufa}
\hrefCMSnoop {} {{ CMS} Collaboration, ``{Observation of a new boson at a mass
  of 125 GeV with the CMS experiment at the LHC}'',} \textit{ Phys.Lett.}
  \textbf{ B716} (2012) 30--61,
\href{http://www.arXiv.org/abs/1207.7235}{\texttt{ arXiv:1207.7235}}.

\bibitem{Kane:1993td}
\hrefCMSnoop {} {G.~L. Kane, C.~F. Kolda, L.~Roszkowski{ et~al.}, ``{Study of
  constrained minimal supersymmetry}'',} \textit{ Phys.Rev.} \textbf{ D49}
  (1994) 6173--6210,
\href{http://www.arXiv.org/abs/hep-ph/9312272}{\texttt{ arXiv:hep-ph/9312272}}.

\bibitem{Beskidt:2012sk}
\hrefCMSnoop {} {C.~Beskidt, W.~de~Boer, D.~Kazakov{ et~al.}, ``{Constraints on
  Supersymmetry from LHC data on SUSY searches and Higgs bosons combined with
  cosmology and direct dark matter searches}'',} \textit{ Eur.Phys.J.} \textbf{
  C72} (2012) 2166,
\href{http://www.arXiv.org/abs/1207.3185}{\texttt{ arXiv:1207.3185}}.

\bibitem{Buchmueller:2013rsa}
\hrefCMSnoop {} {O.~Buchmueller, R.~Cavanaugh, A.~De~Roeck{ et~al.}, ``{The
  CMSSM and NUHM1 after LHC Run 1}'',}
\href{http://www.arXiv.org/abs/1312.5250}{\texttt{ arXiv:1312.5250}}.

\bibitem{Fowlie:2012im}
\hrefCMSnoop {} {A.~Fowlie, M.~Kazana, K.~Kowalska{ et~al.}, ``{The CMSSM
  Favoring New Territories: The Impact of New LHC Limits and a 125 GeV
  Higgs}'',} \textit{ Phys.Rev.} \textbf{ D86} (2012) 075010,
\href{http://www.arXiv.org/abs/1206.0264}{\texttt{ arXiv:1206.0264}}.

\bibitem{Bechtle:2013mda}
\hrefCMSnoop {} {P.~Bechtle, K.~Desch, H.~K. Dreiner{ et~al.}, ``{Constrained
  Supersymmetry after the Higgs Boson Discovery: A global analysis with
  Fittino}'',}
\href{http://www.arXiv.org/abs/1310.3045}{\texttt{ arXiv:1310.3045}}.

\bibitem{Ellwanger:2009dp}
\hrefCMSnoop {} {U.~Ellwanger, C.~Hugonie, and A.~M. Teixeira, ``{The
  Next-to-Minimal Supersymmetric Standard Model}'',} \textit{ Phys.Rept.}
  \textbf{ 496} (2010) 1--77,
\href{http://www.arXiv.org/abs/0910.1785}{\texttt{ arXiv:0910.1785}}.

\bibitem{King:2012tr}
\hrefCMSnoop {} {S.~F. King, M.~M{\"u}hlleitner, R.~Nevzorov{ et~al.},
  ``{Natural NMSSM Higgs Bosons}'',} \textit{ Nucl.Phys.} \textbf{ B870} (2013)
  323--352,
\href{http://www.arXiv.org/abs/1211.5074}{\texttt{ arXiv:1211.5074}}.

\bibitem{Cao:2012fz}
\hrefCMSnoop {} {J.-J. Cao, Z.-X. Heng, J.~M. Yang{ et~al.}, ``{A SM-like Higgs
  near 125 GeV in low energy SUSY: a comparative study for MSSM and NMSSM}'',}
  \textit{ JHEP} \textbf{ 1203} (2012) 086,
\href{http://www.arXiv.org/abs/1202.5821}{\texttt{ arXiv:1202.5821}}.

\bibitem{Belanger:2012tt}
\hrefCMSnoop {} {G.~Belanger, U.~Ellwanger, J.~F. Gunion{ et~al.}, ``{Higgs
  Bosons at 98 and 125 GeV at LEP and the LHC}'',} \textit{ JHEP} \textbf{
  1301} (2013) 069,
\href{http://www.arXiv.org/abs/1210.1976}{\texttt{ arXiv:1210.1976}}.

\bibitem{Ellwanger:2012ke}
\hrefCMSnoop {} {U.~Ellwanger and C.~Hugonie, ``{Higgs bosons near 125 GeV in
  the NMSSM with constraints at the GUT scale}'',} \textit{ Adv.High Energy
  Phys.} \textbf{ 2012} (2012) 625389,
\href{http://www.arXiv.org/abs/1203.5048}{\texttt{ arXiv:1203.5048}}.

\bibitem{King:2012is}
\hrefCMSnoop {} {S.~King, M.~M{\"u}hlleitner, and R.~Nevzorov, ``{NMSSM Higgs
  Benchmarks Near 125 GeV}'',} \textit{ Nucl.Phys.} \textbf{ B860} (2012)
  207--244,
\href{http://www.arXiv.org/abs/1201.2671}{\texttt{ arXiv:1201.2671}}.

\bibitem{Vasquez:2012hn}
\hrefCMSnoop {} {D.~A. Vasquez, G.~Belanger, C.~Boehm{ et~al.}, ``{The 125 GeV
  Higgs in the NMSSM in light of LHC results and astrophysics constraints}'',}
  \textit{ Phys.Rev.} \textbf{ D86} (2012) 035023,
\href{http://www.arXiv.org/abs/1203.3446}{\texttt{ arXiv:1203.3446}}.

\bibitem{Beskidt:2013gia}
\hrefCMSnoop {} {C.~Beskidt, W.~de~Boer, and D.~Kazakov, ``{A comparison of the
  Higgs sectors of the CMSSM and NMSSM for a 126 GeV Higgs boson}'',} \textit{
  Phys.Lett.} \textbf{ B726} (2013) 758--766,
\href{http://www.arXiv.org/abs/1308.1333}{\texttt{ arXiv:1308.1333}}.

\bibitem{Badziak:2013bda}
\hrefCMSnoop {} {M.~Badziak, M.~Olechowski, and S.~Pokorski, ``{New Regions in
  the NMSSM with a 125 GeV Higgs}'',} \textit{ JHEP} \textbf{ 1306} (2013) 043,
\href{http://www.arXiv.org/abs/1304.5437}{\texttt{ arXiv:1304.5437}}.

\bibitem{Das:2011dg}
\hrefCMSnoop {} {D.~Das, U.~Ellwanger, and A.~M. Teixeira, ``{NMSDECAY: A
  Fortran Code for Supersymmetric Particle Decays in the Next-to-Minimal
  Supersymmetric Standard Model}'',} \textit{ Comput.Phys.Commun.} \textbf{
  183} (2012) 774--779,
\href{http://www.arXiv.org/abs/1106.5633}{\texttt{ arXiv:1106.5633}}.

\bibitem{Miller:2003ay}
\hrefCMSnoop {} {D.~Miller, R.~Nevzorov, and P.~Zerwas, ``{The Higgs sector of
  the next-to-minimal supersymmetric standard model}'',} \textit{ Nucl.Phys.}
  \textbf{ B681} (2004) 3--30,
\href{http://www.arXiv.org/abs/hep-ph/0304049}{\texttt{ arXiv:hep-ph/0304049}}.

\bibitem{Staub:2010ty}
\hrefCMSnoop {} {F.~Staub, W.~Porod, and B.~Herrmann, ``{The Electroweak sector
  of the NMSSM at the one-loop level}'',} \textit{ JHEP} \textbf{ 1010} (2010)
  040,
\href{http://www.arXiv.org/abs/1007.4049}{\texttt{ arXiv:1007.4049}}.

\bibitem{Arganda:2012qp}
\hrefCMSnoop {} {E.~Arganda, J.~L. Diaz-Cruz, and A.~Szynkman, ``{Decays of
  $H^0/A^0$ in supersymmetric scenarios with heavy sfermions}'',} \textit{ Eur.
  Phys. J.} \textbf{ C73} (2013), no.~4, 2384,
\href{http://www.arXiv.org/abs/1211.0163}{\texttt{ arXiv:1211.0163}}.

\bibitem{Arganda:2013ve}
\hrefCMSnoop {} {E.~Arganda, J.~Lorenzo Diaz-Cruz, and A.~Szynkman, ``{Slim
  SUSY}'',} \textit{ Phys. Lett.} \textbf{ B722} (2013) 100--106,
\href{http://www.arXiv.org/abs/1301.0708}{\texttt{ arXiv:1301.0708}}.

\bibitem{James:1975dr}
\hrefCMSnoop {} {F.~James and M.~Roos, ``{Minuit: A System for Function
  Minimization and Analysis of the Parameter Errors and Correlations}'',}
  \textit{ Comput.Phys.Commun.} \textbf{ 10} (1975) 343--367.

\bibitem{Beskidt:2014kon}
C.~Beskidt, ``{Supersymmetry in the Light of Dark Matter and a 125 GeV Higgs
  Boson}''.
\newblock PhD thesis, KIT, Karlsruhe, EKP,
2014.
\newblock

\bibitem{Belanger:2010pz}
\hrefCMSnoop {} {G.~Belanger, F.~Boudjema, A.~Pukhov{ et~al.}, ``{micrOMEGAs: A
  Tool for dark matter studies}'',}
  \href{http://www.arXiv.org/abs/1005.4133}{\texttt{ arXiv:1005.4133}}.

\bibitem{Beskidt:2010va}
\hrefCMSnoop {} {C.~Beskidt, W.~de~Boer, T.~Hanisch{ et~al.}, ``{Constraints on
  Supersymmetry from Relic Density compared with future Higgs Searches at the
  LHC}'',} \textit{ Phys.Lett.} \textbf{ B695} (2011) 143--148,
\href{http://www.arXiv.org/abs/1008.2150}{\texttt{ arXiv:1008.2150}}.

\bibitem{Mitani:2015aga}
\hrefCMSnoop {} {{ ATLAS} Collaboration, ``{Search for neutral MSSM Higgs
  Bosons in the h/A/H to $\tau^+ \tau^-$ Decay Mode with the ATLAS
  Detector}'',} \textit{ Nucl. Phys. Proc. Suppl.} \textbf{ 253-255} (2014)
220--221.

\bibitem{Khachatryan:2014wca}
\hrefCMSnoop {} {{ CMS} Collaboration, ``{Search for neutral MSSM Higgs bosons
  decaying to a pair of tau leptons in pp collisions}'',} \textit{ JHEP}
  \textbf{ 10} (2014) 160,
\href{http://www.arXiv.org/abs/1408.3316}{\texttt{ arXiv:1408.3316}}.

\bibitem{Beskidt:2011qf}
\hrefCMSnoop {} {C.~Beskidt, W.~de~Boer, D.~Kazakov{ et~al.}, ``{Constraints
  from the decay $B_s^0 \rightarrow; \mu^+ \mu^-$ and LHC limits on
  Supersymmetry}'',} \textit{ Phys.Lett.} \textbf{ B705} (2011) 493--497,
\href{http://www.arXiv.org/abs/1109.6775}{\texttt{ arXiv:1109.6775}}.

\bibitem{Djouadi:2006bz}
\hrefCMSnoop {} {A.~Djouadi, M.~M. Muhlleitner, and M.~Spira, ``{Decays of
  supersymmetric particles: The Program SUSY-HIT
  (SUspect-SdecaY-Hdecay-InTerface)}'',} \textit{ Acta Phys. Polon.} \textbf{
  B38} (2007) 635--644,
\href{http://www.arXiv.org/abs/hep-ph/0609292}{\texttt{ arXiv:hep-ph/0609292}}.

\bibitem{Aad:2011rv}
\hrefCMSnoop {} {{ ATLAS} Collaboration, ``{Search for neutral MSSM Higgs
  bosons decaying to tau$^+$ tau$^-$ pairs in proton-proton collisions at
  sqrt(s) = 7 TeV with the ATLAS detector}'',} \textit{ Phys.Lett.} \textbf{
  B705} (2011) 174--192,
\href{http://www.arXiv.org/abs/1107.5003}{\texttt{ arXiv:1107.5003}}.

\bibitem{Harlander:2012pb}
\hrefCMSnoop {} {R.~V. Harlander, S.~Liebler, and H.~Mantler, ``{SusHi: A
  program for the calculation of Higgs production in gluon fusion and
  bottom-quark annihilation in the Standard Model and the MSSM}'',} \textit{
  Comput. Phys. Commun.} \textbf{ 184} (2013) 1605--1617,
\href{http://www.arXiv.org/abs/1212.3249}{\texttt{ arXiv:1212.3249}}.

\bibitem{Harlander:2002wh}
\hrefCMSnoop {} {R.~V. Harlander and W.~B. Kilgore, ``{Next-to-next-to-leading
  order Higgs production at hadron colliders}'',} \textit{ Phys. Rev. Lett.}
  \textbf{ 88} (2002) 201801,
\href{http://www.arXiv.org/abs/hep-ph/0201206}{\texttt{ arXiv:hep-ph/0201206}}.

\bibitem{Harlander:2003ai}
\hrefCMSnoop {} {R.~V. Harlander and W.~B. Kilgore, ``{Higgs boson production
  in bottom quark fusion at next-to-next-to leading order}'',} \textit{ Phys.
  Rev.} \textbf{ D68} (2003) 013001,
\href{http://www.arXiv.org/abs/hep-ph/0304035}{\texttt{ arXiv:hep-ph/0304035}}.

\bibitem{Aglietti:2004nj}
\hrefCMSnoop {} {U.~Aglietti, R.~Bonciani, G.~Degrassi{ et~al.}, ``{Two loop
  light fermion contribution to Higgs production and decays}'',} \textit{ Phys.
  Lett.} \textbf{ B595} (2004) 432--441,
\href{http://www.arXiv.org/abs/hep-ph/0404071}{\texttt{ arXiv:hep-ph/0404071}}.

\bibitem{Bonciani:2010ms}
\hrefCMSnoop {} {R.~Bonciani, G.~Degrassi, and A.~Vicini, ``{On the Generalized
  Harmonic Polylogarithms of One Complex Variable}'',} \textit{ Comput. Phys.
  Commun.} \textbf{ 182} (2011) 1253--1264,
\href{http://www.arXiv.org/abs/1007.1891}{\texttt{ arXiv:1007.1891}}.

\bibitem{Degrassi:2010eu}
\hrefCMSnoop {} {G.~Degrassi and P.~Slavich, ``{NLO QCD bottom corrections to
  Higgs boson production in the MSSM}'',} \textit{ JHEP} \textbf{ 11} (2010)
  044,
\href{http://www.arXiv.org/abs/1007.3465}{\texttt{ arXiv:1007.3465}}.

\bibitem{Degrassi:2011vq}
\hrefCMSnoop {} {G.~Degrassi, S.~Di~Vita, and P.~Slavich, ``{NLO QCD
  corrections to pseudoscalar Higgs production in the MSSM}'',} \textit{ JHEP}
  \textbf{ 08} (2011) 128,
\href{http://www.arXiv.org/abs/1107.0914}{\texttt{ arXiv:1107.0914}}.

\bibitem{Degrassi:2012vt}
\hrefCMSnoop {} {G.~Degrassi, S.~Di~Vita, and P.~Slavich, ``{On the NLO QCD
  Corrections to the Production of the Heaviest Neutral Higgs Scalar in the
  MSSM}'',} \textit{ Eur. Phys. J.} \textbf{ C72} (2012) 2032,
\href{http://www.arXiv.org/abs/1204.1016}{\texttt{ arXiv:1204.1016}}.

\bibitem{Liebler:2015bka}
\hrefCMSnoop {} {S.~Liebler, ``{Neutral Higgs production at proton colliders in
  the CP-conserving NMSSM}'',} \textit{ Eur. Phys. J.} \textbf{ C75} (2015),
  no.~5, 210,
\href{http://www.arXiv.org/abs/1502.07972}{\texttt{ arXiv:1502.07972}}.

\bibitem{Hahn:2013ria}
\hrefCMSnoop {} {T.~Hahn, S.~Heinemeyer, W.~Hollik{ et~al.}, ``{High-Precision
  Predictions for the Light CP -Even Higgs Boson Mass of the Minimal
  Supersymmetric Standard Model}'',} \textit{ Phys. Rev. Lett.} \textbf{ 112}
  (2014), no.~14, 141801,
\href{http://www.arXiv.org/abs/1312.4937}{\texttt{ arXiv:1312.4937}}.

\bibitem{Frank:2006yh}
\hrefCMSnoop {} {M.~Frank, T.~Hahn, S.~Heinemeyer{ et~al.}, ``{The Higgs Boson
  Masses and Mixings of the Complex MSSM in the Feynman-Diagrammatic
  Approach}'',} \textit{ JHEP} \textbf{ 02} (2007) 047,
\href{http://www.arXiv.org/abs/hep-ph/0611326}{\texttt{ arXiv:hep-ph/0611326}}.

\bibitem{Degrassi:2002fi}
\hrefCMSnoop {} {G.~Degrassi, S.~Heinemeyer, W.~Hollik{ et~al.}, ``{Towards
  high precision predictions for the MSSM Higgs sector}'',} \textit{ Eur. Phys.
  J.} \textbf{ C28} (2003) 133--143,
\href{http://www.arXiv.org/abs/hep-ph/0212020}{\texttt{ arXiv:hep-ph/0212020}}.

\bibitem{Heinemeyer:1998np}
\hrefCMSnoop {} {S.~Heinemeyer, W.~Hollik, and G.~Weiglein, ``{The Masses of
  the neutral CP - even Higgs bosons in the MSSM: Accurate analysis at the two
  loop level}'',} \textit{ Eur. Phys. J.} \textbf{ C9} (1999) 343--366,
\href{http://www.arXiv.org/abs/hep-ph/9812472}{\texttt{ arXiv:hep-ph/9812472}}.

\bibitem{Heinemeyer:1998yj}
\hrefCMSnoop {} {S.~Heinemeyer, W.~Hollik, and G.~Weiglein, ``{FeynHiggs: A
  Program for the calculation of the masses of the neutral CP even Higgs bosons
  in the MSSM}'',} \textit{ Comput. Phys. Commun.} \textbf{ 124} (2000) 76--89,
\href{http://www.arXiv.org/abs/hep-ph/9812320}{\texttt{ arXiv:hep-ph/9812320}}.

\bibitem{CMS-PAS-EXO-15-004}
\href {https://cds.cern.ch/record/2114808} {{ CMS} Collaboration, ``{Search for
  new physics in high mass diphoton events in proton-proton collisions at
  $\sqrt{s} = 13$ TeV}'',} Technical Report CMS-PAS-EXO-15-004, CERN, Geneva,
  2015.

\bibitem{ATLAS-CONF-2015-081}
\href {http://cds.cern.ch/record/2114853} {{ ATLAS} Collaboration, ``{Search
  for resonances decaying to photon pairs in 3.2 fb$^{-1}$ of $pp$ collisions
  at $\sqrt{s}$ = 13 TeV with the ATLAS detector}'',} Technical Report
  ATLAS-CONF-2015-081, CERN, Geneva, Dec, 2015.

\bibitem{Ellis:2015oso}
\hrefCMSnoop {} {J.~Ellis, S.~A.~R. Ellis, J.~Quevillon{ et~al.}, ``{On the
  Interpretation of a Possible $\sim 750$ GeV Particle Decaying into $\gamma
  \gamma$}'',} \textit{ JHEP} \textbf{ 03} (2016) 176,
\href{http://www.arXiv.org/abs/1512.05327}{\texttt{ arXiv:1512.05327}}.

\bibitem{King:2016wep}
\hrefCMSnoop {} {S.~F. King and R.~Nevzorov, ``{750 GeV Diphoton Resonance from
  Singlets in an Exceptional Supersymmetric Standard Model}'',} \textit{ JHEP}
  \textbf{ 03} (2016) 139,
\href{http://www.arXiv.org/abs/1601.07242}{\texttt{ arXiv:1601.07242}}.

\bibitem{Petersson:2015mkr}
\hrefCMSnoop {} {C.~Petersson and R.~Torre, ``{The 750 GeV diphoton excess from
  the goldstino superpartner}'',}
\href{http://www.arXiv.org/abs/1512.05333}{\texttt{ arXiv:1512.05333}}.

\bibitem{Bhattacharya:2016lyg}
\hrefCMSnoop {} {S.~Bhattacharya, S.~Patra, N.~Sahoo{ et~al.}, ``{750 GeV
  Di-photon excess at CERN LHC from a dark sector assisted scalar decay}'',}
\href{http://www.arXiv.org/abs/1601.01569}{\texttt{ arXiv:1601.01569}}.

\bibitem{Chang:2015sdy}
\hrefCMSnoop {} {J.~Chang, K.~Cheung, and C.-T. Lu, ``{Interpreting the 750 GeV
  Di-photon Resonance using photon-jets in Hidden-Valley-like models}'',}
  \textit{ Phys. Rev.} \textbf{ D93} (2016) 075013,
\href{http://www.arXiv.org/abs/1512.06671}{\texttt{ arXiv:1512.06671}}.

\end{thebibliography}\endgroup








\end{document}